\documentclass[a4paper,11pt]{article}
\usepackage{jcappub}
\usepackage[T1]{fontenc} 

\usepackage{amsmath} 
\def\overbigdot#1{\overset{\hbox{\tiny$\bullet$}}{#1}}

\title{Little ado about everything II:\\
an `emergent' dark energy from structure formation 
to rule cosmic tensions}

\author[a,b,c,d,*]{Andrea Lapi,}
\author[a,b,*]{Balakrishna S. Haridasu,}
\author[e,a,b]{Lumen Boco,}
\author[a,b]{Marcos M. Cueli,}
\author[a,b,c]{Carlo Baccigalupi,}
\author[a,b]{Luigi Danese}

\affiliation[a]{SISSA, Via Bonomea 265, 34136 Trieste, Italy}
\affiliation[b]{Institute for Fundamental Physics of the Universe (IFPU), Via Beirut 2, 34014 Trieste, Italy}
\affiliation[c]{INFN-Sezione di Trieste, via Valerio 2, 34127 Trieste, Italy}
\affiliation[d]{IRA-INAF, Via Gobetti 101, 40129 Bologna, Italy}
\affiliation[e]{Universitat Heidelberg, Zentrum fur Astronomie, Institut fur theoretische Astrophysik, Albert-Ueberle-Str. 3, 69120 Heidelberg, Germany}
\affiliation[*]{corresponding authors}

\emailAdd{lapi@sissa.it, sharidas@sissa.it}

\abstract{The $\eta$CDM framework by \cite{Lapi2023} is a new cosmological model aimed to cure some drawbacks of the standard $\Lambda$CDM scenario, such as the origin of the accelerated expansion at late times, the cosmic tensions, and the violation of the cosmological principle due to the progressive development of inhomogeneous/anisotropic conditions in the Universe during structure formation. To this purpose, the model adopts a statistical perspective envisaging a stochastic evolution of large-scale patches in the Universe with typical sizes $10-50\, h^{-1}$ Mpc, which is meant to describe the complex gravitational processes leading to the formation of the cosmic web. The stochasticity among different patches is technically rendered via the diverse realizations of a multiplicative noise term (`a little ado') in the cosmological equations, and the overall background evolution of the Universe is then operationally defined as an average over the patch ensemble. In this paper we show that such an ensemble-averaged evolution in $\eta$CDM can be described in terms of a spatially flat cosmology and of an `emergent' dark energy with a time-dependent equation of state, able to originate the cosmic acceleration with the right timing and to solve the coincidence problem. Moreover, we provide a cosmographic study of the $\eta$CDM model, suitable for quick implementation in the analysis of future observations. Then we test the $\eta$CDM model against the most recent supernova type-I$a$, baryon acoustic oscillations and structure growth rate datasets, finding an excellent agreement. Remarkably, we demonstrate that $\eta$CDM is able to alleviate simultaneously both the $H_0$ and the $f\sigma_8$ tensions. Finally, we discuss that the Linders' diagnostic test could be helpful to better distinguish $\eta$CDM from the standard scenario in the near future via upcoming galaxy redshift surveys at intermediate redshifts such as those being conducted by the \texttt{Euclid} mission.}

\keywords{Cosmology -- structure formation -- dark energy}

\arxivnumber{2502.05823}

\begin{document}
\maketitle
\flushbottom

\section{Introduction}\label{sec|intro}

Once upon a time, there was a standard cosmological framework, dubbed $\Lambda$CDM, written in stone. In fact, such a model proved extremely successful in the era of precision cosmology, by efficiently reproducing a large number of observations related to the cosmic microwave background (CMB; e.g, \cite{Bennett2013,Aiola2020,SPT2021,Planck2020}), type-I$a$ supernovae (SN; e.g., \cite{Perlmutter1999,Scolnic2018,Brout2022,DES2024}), baryon acoustic oscillations (BAO; e.g., \cite{Aubourg2015,Alam2017,Alam2021,Adame2025}), structure growth rates (e.g., \cite{Blake2012,Beutler2012,Okumura2016,Pezzotta2017,Said2020,Turner2022}), galaxy cluster counts (e.g., \cite{White1993,Mantz2022,Ghirardini2024}), and many others. However, in recent years the $\Lambda$CDM paradigm has been increasingly challenged by new observations and state-of-the-art numerical/theoretical developments. 

There are three main, well-recognized downsides of the standard cosmological framework. First, it postulates the existence of a mysterious dark energy component violating the strong energy condition of general relativity, that triggers the accelerated expansion of the Universe at late times (as revealed by type-I$a$ SN \cite{Riess98, Scolnic2022} and supported by BAOs) and enforces a closely flat spatial geometry (as required by CMB and BAO data). The observed value of the present-day dark energy density appears weird, being far below any natural scale in particle physics but surprisingly similar to the current matter density (aka the `coincidence' problem; e.g., \cite{Zeldovich1968,Weinberg1989}). Recent data from \texttt{DES} and \texttt{DESI} \cite{DES2024,Adame2025}, though still debated \cite{Colgain2024a,Colgain2024b,Sousa2025}, possibly even suggest some form of dynamical dark energy with an evolving equation of state, which may call for an exotic extension of the basic $\Lambda$CDM framework.

Second, the $\Lambda$CDM model struggles in solving a few cosmic tensions: the $H_0$ tension concerns the background evolution of the Universe, and highlights a disagreement between the determination of the Hubble constant from late-time observables and the CMB (e.g., \cite{Riess2016,Riess2022,Riess2024}); the $f\sigma_8$ and $S_8\equiv \sigma_8\,(\Omega_m/0.3)^{1/2}$ tensions signal a deficit in these combinations of cosmological parameters ($\sigma_8$ being the rms amplitude of the matter power spectrum smoothed on a scale of $8\, h^{-1}$ Mpc, $\Omega_m$ being the matter density, and $f$ being the growth rate of linear perturbations) as constrained from galaxy redshift surveys and cosmic shear data with respect to the CMB expectations (e.g., \cite{Heymans2013,DiValentino2021b,Secco2022}). Of these the $H_0$ and the $f\sigma_8$ tensions are considered the most serious, since the $S_8$ tension may be possibly mitigated or explained by baryonic effects (e.g., some form of feedback from astrophysical sources, see \cite{AmonEfst2022}).

Third, the $\Lambda$CDM model is deeply rooted on the assumption of a homogeneous and isotropic Universe, that allows to adopt a global Friedman-Robertson-Walker metric and to assign a spatially-independent value of the cosmological quantities at any given cosmic time. However, recent data suggest that such an assumption may be questioned at late times on scales $\lesssim 50\, h^{-1}$ Mpc (and likely on even larger ones, see \cite{Aluri2023,Courtois2025}), that are crucial for several cosmological observables, as well as for the calibration of the cosmic ladder (see \cite{Pesce2020,Boruah2020,Riess2022,Tully2023}). Numerical simulations indicate that such violations of isotropy and homogeneity are induced by the complex gravitational processes leading to the formation of the cosmic web, a spider-like network of quasi-linear structures encompassing knots, filaments, walls and voids over a wide range of sizes $R\sim 10-50\, h^{-1}$ Mpc  (e.g., \cite{Libeskind2018,Wilding2021,Douglass2023}). Matter flows along the cosmic web structures cause the Universe to be increasingly dominated in volume by voids, and the distribution of overdensities smoothed over scales $\lesssim 50\, h^{-1}$ Mpc shifts for $z\lesssim 1$ from the initial Gaussian shape to a lognormal one, as expected theoretically (e.g., \cite{Coles1991,Repp2017,Repp2018}), and inferred observationally from reconstruction of the large-scale velocity field (e.g., \cite{Tully2008,Tully2023,Hoffman2024,Courtois2025}). 

It is worth mentioning that there are additional open issues, like the nature of DM and the physics of inflation, but these ingredients of the model subtend more successes than doubts (for reviews, see \cite{Bertone2018,Ellis2023}). As to the former, despite the absence for a firm detection of DM particles in colliders or with direct and indirect searches in the sky, the cold DM paradigm (envisaging DM to be constituted by weakly interacting particles that are already non-relativistic at decoupling) explains extremely well the formation of cosmic structures as observed across the history of the Universe. As to the latter, although inflation is still a not perfectly understood physical process, it can solve some fundamental issues like the flatness and the horizon problems, and can provide a natural perspective for understanding the quantum origin of a nearly scale-invariant spectrum of primordial fluctuations. Therefore the standard lore is that the challenge is more on the experimental than on the foundational side (see \cite{Peebles2020,Efstathiou2023}). 

The aforementioned pressing issues, and in particular the origin of the cosmic acceleration, have called for a revision of the standard theory, stimulating the formulation of several cosmological models alternative to it. A non-exhaustive list includes: ab-initio modified gravity theories with additional degrees of freedom in the matter and/or gravitational action (e.g., \cite{Clifton2012,Nojiri2017,Saridakis2021}); phenomenological modifications of the Friedmann equations by non-linear terms in the matter density (Cardassian scenarios; e.g., \cite{Freese2002,Xu2012,Magana2018}); alteration of the mass-energy conservation equations with bulk viscosity terms (e.g., \cite{Lima1988,Brevik2011,Herrera2020}); biases in cosmological parameter estimates due to local underdensities or to the preferential association of standard candles with overdense regions (e.g., \cite{Celerier2000,Alnes2006,Deledicque2023}); deterministic backreaction models highlighting the possible impact of small-scale anisotropies/inhomogeneities in the matter field on the overall cosmic expansion rate (e.g., \cite{Buchert1997,Buchert2000,Buchert2000b,Buchert2008,Buchert2012,Buchert2013,Rasanen2010,Racz2017,Koksbang2020,Schander2021}) or stressing the role of voids in providing gravitational energy gradients and variance in clock rates that can be misinterpreted as an apparent dark energy (e.g., the `timescape' model, see \cite{Wiltshire2007,Wiltshire2009,Seifert2025}). In addition, recent studies in numerical relativity have pointed out that sampling effects associated to the inhomogeneous/anisotropic nature of structure formation could have an appreciable impact on the inference of cosmological observables \cite{Koksbang2024,Macpherson2024}.

Yet no definite paradigm shift has took place, because these alternative theories, despite curing some critical aspects of $\Lambda$CDM, are generally more complex (i.e., characterized by more parameters) and do not allow a systematic and coherent performance testing against the exceedingly large amount of available cosmological data (as instead $\Lambda$CDM does). As a matter of fact, the excellent performance of the $\Lambda$CDM model in coherently reproducing cosmological observations over an extremely wide range of spatial and temporal scales may suggest that any alternative to it should be rather close in terms of phenomenological outcomes, like for example the timing and the strength of the cosmic acceleration, the nearly flat geometry of the Universe, etc. (see \cite{Peebles2020,Turner2022,Efstathiou2023}); therefore it would be non trivial to perform model selection if not basing on specifically designed diagnostics or convincing theoretical arguments.

Following these lines, in \cite{Lapi2023} we have proposed a novel model of the Universe dubbed $\eta$CDM that is aimed to cure the aforementioned drawbacks of the standard scenario, at the same time allowing for an extensive testing against several cosmological observables and laying a new theoretical perspective concerning the very foundations of the cosmological landscape. In fact, the $\eta$CDM model drops the deterministic viewpoint underlying the standard `cosmological principle' (i.e., the statement that the distribution and motion of matter in any sufficiently large spatial region of the Universe are much the same in any other region) and rather adopts a statistical perspective envisaging the evolution of large-scale patches of the Universe as a spatially-stationary stochastic process\footnote{The mathematical framework of stochastic processes is nowadays largely exploited by scientific communities interested in complex systems, especially to describe the formation of spatial and/or temporal structures in physics, chemistry, biology and many other fields (see \cite{Risken1996,Paul2013}). In cosmology, the theory of stochastic processes has been applied to investigate inflation (e.g., \cite{Vilenkin1983,Salopek1991,Cruces2022}) and to predict the DM halo mass function and related statistics
(e.g., \cite{Bond1991,Mo1996,Lapi2020,Lapi2022}).} (i.e., meaning that the probability of the happenings that lead to structure formation in the Universe on sufficiently large scales are similar at every location and in any direction in space). This change of foundational basis was pioneered by \cite{Neyman1962} and recently advocated by \cite{Peebles2022} to constitute a possibly crucial step forward in formulating an improved cosmological model of the Universe. 

As a consequence, in $\eta$CDM the spatially-averaged evolution of patches with sizes $10-50\, h^{-1}$ Mpc includes a degree of scale-dependent stochasticity. Technically, the evolution of the different patches as a function of cosmic time is rendered via the diverse realizations of a multiplicative noise term (a `little ado') in the mass-energy evolution equation. At any given cosmic time, sampling the ensemble of patches generates a non-trivial distribution of the cosmological quantities; the shape of the noise term is designed in such a way that the overdensity field smoothed on scales $\lesssim 50\, h^{-1}$ Mpc features a closely lognormal distribution, as observed at late times (see references above). The overall behavior of the Universe is then operationally defined as an average over the patch ensemble. Tuning the noise against a wealth of cosmological datasets, in \cite{Lapi2023} we were able to show that in the $\eta$CDM model an accelerated expansion of the Universe is enforced at late cosmic times, without the need for any postulated dark energy component, while substantially relieving the $H_0$ tension. The acceleration of the ensemble-averaged evolution originates because, as structure formation proceeds in the Universe, an increasing amount of its volume is statistically occupied by low-density regions (i.e., voids) featuring an enhanced expansion rate (as confirmed by general relativity simulations, see \cite{Williams2025}). 

In the present work, we further test the $\eta$CDM model against the most recent type-I$a$ SN, BAO and structure growth rate datasets (including \texttt{Pantheon+} and \texttt{DESI} samples) to demonstrate that such a framework is able to alleviate simultaneously both the $H_0$ and the $f\sigma_8$ tensions. We also offer a new look to the $\eta$CDM model, showing that its ensemble-averaged evolution can be rendered in terms of a spatially flat cosmology and of an emergent dark energy component with a time-dependent equation of state. In addition, we also provide a new cosmographic analysis of the $\eta$CDM model, suitable for quick implementation in the analysis of future datasets. 

The plan of the paper is as follows. In Section \ref{sec|theory} we briefly recap the basics of the $\eta$CDM model and provide the aforementioned novel theoretical viewpoint. In Section \ref{sec|data} we describe the datasets, detail our analysis and present our results. In Section \ref{sec|discussion} we discuss our main findings. In Section \ref{sec|summary} we summarize the present work, drawing general conclusions and outlining future perspectives.
 
\section{A primer of the \texorpdfstring{$\eta$CDM}{etaCDM} model}\label{sec|theory}

In this Section we first briefly recap the basics of the $\eta$CDM model as presented in \cite{Lapi2023}, and then provide a novel look at it, showing that the ensemble-averaged evolution of the Universe in such a framework can be rendered as a spatially flat cosmology equipped with an `emergent' dark energy component.

\subsection{Basics}\label{sec|basics}

The foundational change of paradigm at the hearth of the $\eta$CDM model is to drop the classic deterministic formulation of the cosmological principle, and to adopt instead a statistical perspective 
envisaging the evolution of large-scale patches of the Universe as a spatially-stationary stochastic process. In other words, when smoothed on scales $10-50\, h^{-1}$ Mpc pertaining to the formation of the cosmic web, cosmological quantities are not assumed to take the same value at any location (dependent only on time, like in the standard theory), but rather to take different values at different locations according to well-definite probability distributions (with these  dependent on time but independent of position and or orientation). This view is largely supported by numerical simulations and observations of the smoothed density field (see references in Section \ref{sec|intro}; also Appendix \ref{app|spave}). In fact, the latter's distribution follows an approximately Gaussian shape at early times reflecting the initial conditions, and becomes progressively lognormal at late times; this is due to the complex gravitational processes associated to structure formation which tend to aggregate matter in the filaments/nodes of the cosmic web, while subtracting it from other regions that will become the voids. 

Thus in $\eta$CDM the evolution of different large-scale patches in the Universe with sizes $10-50\, h^{-1}$ Mpc includes a degree of stochasticity, which is rendered via the diverse realizations of a multiplicative noise term in the cosmological equations, promoting these to a stochastic differential system
\begin{equation}\label{eq|basics}
\left\{
\begin{aligned}
H^2 &= \frac{8\pi G}{3}\,(\rho_m+\rho_\gamma) - \frac{k}{a^2}~,\\
&\\
\dot\rho_{m,\gamma} &= -3 H\, \rho_{m,\gamma}\,(1+\omega_{m,\gamma}) + \zeta\, \rho_{m,\gamma}\, H^{\alpha}\, H_\star^{-\alpha+1/2}\, \eta(t)~;\\
\end{aligned}
\right.
\end{equation}
here $a$ is the scale factor, $H\equiv \dot a/a$ the Hubble rate, $k$ the curvature constant, $\rho_{m,\gamma}$ the matter and radiation densities with $(\omega_m,\omega_\gamma)=(0,1/3)$; moreover, $\eta(t)$ is a Gaussian white noise of the Stratonovich type with ensemble-average statistical properties $\overline{\eta(t)}=0$ and $\overline{\eta(t)\eta(t')}=2\, \delta_{\rm D}(t-t')$, $\zeta>0$ and $\alpha<0$ are two parameters regulating the strength and redshift dependence of the noise\footnote{Note that in principle the noise terms in the matter and radiation components could feature different shapes (e.g., different parameters $\zeta$ and $\alpha$), but here for the sake of simplicity we have assumed similar expressions for them, though each one is proportional to the corresponding energy density $\rho_{m,\gamma}$. In fact, this proportionality makes any specific assumption on the noise term in the radiation component irrelevant since at late times when noise becomes important $\rho_\gamma$ is negligible, while toward high redshift when $\rho_\gamma$ becomes appreciable the noise is suppressed (since $\alpha<0$ and $H$ grows).}, and $H_\star=100$ km s$^{-1}$ Mpc$^{-1}$ is a reference value of the Hubble rate that is present for dimensional consistency (since $\eta$ has dimension of one over square root of time). We stress that the above equations refer to patches of the Universe with a given size $10-50\, h^{-1}$ Mpc and that the quantities appearing there are spatially-averaged (i.e., smoothed) over such a scale. Thus if we imagine to tessellate the Universe with these patches, each realization of the noise $\eta(t)$ will correspond to a slightly different evolution for each patch. At any given cosmic time, sampling the ensemble of patches will originate a non-trivial distribution of the smoothed cosmological quantities. The evolution of the Universe as a whole is operationally defined as the ensemble-average over the different patches.

In Appendix \ref{app|spave} the form of these equations for the matter component is shown to emerge quite naturally from a spatial-averaging procedure in Newtonian cosmology\footnote{Note that for notational economy in the above equations we have dropped the symbols $\langle\cdot\rangle$ used in the Appendix to denote spatial averaging, but every quantity is meant to be smoothed on a given spatial scale. In the same vein, we have indicated with $\rho$ the density source although in Appendix \ref{app|spave} it is shown that this should be an effective quantity $\rho_{\rm eff}$ including a sub-leading contribution from the noise itself; in our framework $\rho_{\rm eff}\approx \rho$ since the noise describes only small deviation from homogeneity/isotropy.}. Moreover, there it is shown that the strength of the noise $\zeta\propto \langle\theta^2\rangle$ can be related to the spatially-averaged variance of the peculiar velocity field divergence $\theta \equiv \nabla\cdot \bf \dot x$, and that there are good arguments to expect an evolution parameter $\alpha\gtrsim -1$. Given that the stochasticity is associated to structure formation, on very large scales and/or at early times the impact of the noise will be negligible and the average evolution of the Universe will be indistinguishable from the standard $\Lambda$CDM scenario. However, as already mentioned in Section \ref{sec|intro} there are robust evidences from observations and from numerical simulations that appreciable values of $\langle\theta^2\rangle$ applies at late times on scales of several tens Mpc associated to the emergence of the cosmic web, that are actually quite critical for many cosmological data, such as the calibration of the cosmic ladder (e.g., \cite{Tully2023,Hoffman2024}). Thus it is natural to expect that the stochasticity will induce some non-trivial effects on the late-time evolution of the Universe, at least when this is operationally defined as an ensemble-average over different patches with such a size. Note that the impact of adding a phenomenological noise term in the evolution of perturbations due to gravitational instability has been previously considered by \cite{Buchert1999}.

To proceed, it is convenient to introduce the normalized Hubble rate $h\equiv H/H_\star$, the normalized time $\tau\equiv H_\star\, t$ so that $\eta(\tau)=\eta(t)/\sqrt{H_\star}$ applies, and the density parameters $\Omega_{m,\gamma}\equiv 8\pi G\rho_{m,\gamma}/3 H^2$. In this way we can put the equations above in the adimensional form:
\begin{equation}
\left\{
\begin{aligned}
\dot h &= h^2\, \left(-1-\frac{\Omega_m}{2}-\Omega_\gamma\right) + \frac{\zeta}{2}\, (\Omega_m+\Omega_\gamma)\, h^{\alpha+1}\, \eta(\tau)~,\\
&\\
\dot{\Omega}_m &=\Omega_m\, h\, (-1+\Omega_m+2\,\Omega_\gamma)+\zeta\,(1-\Omega_m-\Omega_\gamma)\, \Omega_m\, h^{\alpha}\, \eta(\tau)~,\\
&\\
\dot{\Omega}_\gamma &=\Omega_\gamma\, h\, (-2+\Omega_m+2\,\Omega_\gamma)+\zeta\,(1-\Omega_m-\Omega_\gamma)\, \Omega_\gamma\, h^{\alpha}\, \eta(\tau)~,
\end{aligned}
\right.
\end{equation}
where overdot means now differentiation with respect to $\tau$. In \cite{Lapi2023} we have shown that the evolution of the Universe implied by these stochastic equations includes a dominant ensemble-averaged component common to all the patches, plus a small random component superimposed to it that make the evolution of each patch slightly different from the others. In the present work we are only concerned with the ensemble-averaged evolution. To a first approximation this can be easily derived from the Kramers-Moyal drift and diffusion coefficients associated to the stochastic system (see \cite{Lapi2023} for details). Specifically, for a generic set of equations for the variables $\xi\equiv\{\xi_i\}$ subject to Stratonovich-type noise of the form $\dot \xi_i=f_i({\bf \xi})+g_{ij}({\bf \xi})\, \eta_j(t)$, the ensemble-averaged deterministic components $\{{\bf \bar{\xi}}_i\}$ are approximately given by $\dot{\bar{\xi}}_i = f_i({\bf{\bar \xi}})+g_{kj}\, \partial_k\, g_{ij}(\bf{\bar \xi})$; the second addenda represents a noise-induced drift that come from the multiplicative nature of the stochasticity. In the present context we get:
\begin{equation}\label{eq|friednew}
\left\{
\begin{aligned}
\dot{\bar{h}} &= \bar{h}^2\,\left(-1-\cfrac{\bar\Omega_m}{2}-\bar\Omega_\gamma\right)+\cfrac{\zeta^2}{2}\,(\bar{\Omega}_m+\bar{\Omega}_\gamma)\, \bar{h}^{2\alpha+1}\, \left[1-\cfrac{1-\alpha}{2}\, (\bar\Omega_m+\bar\Omega_\gamma)\right]~,\\
&\\
\dot{\bar{\Omega}}_m &=\bar\Omega_m\, \bar h\, (-1+\bar\Omega_m+2\,\bar\Omega_\gamma)+\zeta^2\,(1-\bar\Omega_m-\bar\Omega_\gamma)\, \bar\Omega_m\, \bar  h^{2\alpha}\, \left[1-\cfrac{4-\alpha}{2}\, 
(\bar\Omega_m+\bar\Omega_\gamma)\right]~,\\
&\\
\dot{\bar{\Omega}}_\gamma &=\bar\Omega_\gamma\, \bar h\, (-2+\bar\Omega_m+2\,\bar\Omega_\gamma)+\zeta^2\,(1-\bar\Omega_m-\bar\Omega_\gamma)\, \bar\Omega_\gamma\, \bar  h^{2\alpha}\, \left[1-\cfrac{4-\alpha}{2}\, (\bar\Omega_m+\bar\Omega_\gamma)\right]~,
\end{aligned}
\right.
\end{equation}
which are the basic equations of the $\eta$CDM model derived in \cite{Lapi2023}. Hereafter the bar over the variables is omitted for notational economy, but we stress that all the quantities are ensemble-averaged. 

\subsection{An `emergent' dark energy}\label{sec|emergentDE}

We now innovate with respect to \cite{Lapi2023} by taking a more convenient and transparent route. First, we combine the above equations to obtain 
\begin{equation}\label{eq|intconstr}                     
2\, \frac{\dot h}{h}+2\,h-\frac{3}{2}\,\zeta^2\,h^{2\alpha}\,(\Omega_m+\Omega_\gamma)^2=\frac{\dot\Omega_m+\dot\Omega_\gamma}{1-\Omega_m-\Omega\gamma}~;
\end{equation}
then we integrate in time to derive the ensemble-averaged analogue of the Friedmann constraint
\begin{equation}\label{eq|constr}
\Omega_\eta \equiv -\frac{\kappa}{h^2\, a^2}\, e^{\frac{3}{2}\zeta^2\, \int_0^\tau{\rm d}\tau\; h^{2\alpha}\, (\Omega_m+\Omega_\gamma)^2}=1-\Omega_m-\Omega_\gamma~,
\end{equation}
where $\kappa$ is a constant of integration and $\Omega_\eta$ is a dark energy-like term `emergent' from the noise. In terms of the latter, the ensemble-average Universe is formally equivalent to a flat cosmology with $\Omega_m+\Omega_\gamma+\Omega_\eta=1$. 

To interpret $\kappa$ in physical terms we may note that in the early Universe the effect of the noise is negligible (the integral in the argument of the exponential vanishes) and the emergent dark energy behaves like a standard curvature component since $\Omega_\eta \simeq -\kappa/h^{2}\,a^2$ applies. In fact, in the closely homogeneous/isotropic primordial patches of the Universe, the constant $k\propto \langle \mathcal{R}\rangle$ appearing in Equations (\ref{eq|basics}) is directly linked to the spatially-averaged scalar curvature $\langle \mathcal{R}\rangle$ of the patch (see Equation \ref{eq|kappaspave}), and thus $\kappa$ could be safely identified at early times with an ensemble-averaged spatial curvature constant. For Equation (\ref{eq|constr}) to be consistent at late times the $\kappa$ is bound to be negative since $1-\Omega_{m,0}-\Omega_{\gamma,0}>0$. This implies that the ensemble-averaged spatial curvature in the early Universe must also be negative, though it can be arbitrarily small (this is allowed since $\kappa/h^2\, a^2$ tends to zero toward the Big Bang when $1-\Omega_m-\Omega_\gamma$ is tiny), so as not to violate cosmological data \cite{Planck2020}; however, if initially it is exactly zero, then $\kappa=0$ so that $\Omega_\eta$ will stay null and no emergent dark energy will be originated, even when the noise term is present.
At late times the direct proportionality between $k$ and the spatially-averaged scalar curvature $\langle\mathcal{R}\rangle$ of the patch is broken, since also inhomogeneities/anisotropies due to structure formation can develop curvature in general relativity (see Equation \ref{eq|kappaspave}), a positive one in the filaments/sheets of the cosmic web and a negative one in the voids. However, when ensemble-averaging over all the patches, it is plausible to expect that the original small overall curvature has not been substantially altered, and that the ensemble-averaged Universe will stay close to flatness. In this perspective of ensemble-averaged evolution, the exponential term in Equation (\ref{eq|constr}) effectively acts in transforming a small curvature term at early epochs into an emergent dark energy at late times.

Nevertheless, other interpretations are possible. For example, in \cite{Lapi2023} we took an alternative viewpoint by writing Equation (\ref{eq|constr}) in the form of a modified Friedman constraint $\Omega_\kappa\equiv -\kappa/h^2\,a^2 = (1-\Omega_m-\Omega_\gamma)\, e^{-3\zeta^2/2\,\int{\rm d}\tau\, h^{2\alpha}\, (\Omega_m+\Omega_\gamma)^2}$, so that one can interpret $\Omega_\kappa$ as an emergent (negative) curvature-like component with a non-standard evolution. To fit cosmological data, in \cite{Lapi2023} we found that $\Omega_k$ is bound to be small in the remote past, to grow at a maximal value $\Omega_k\lesssim 0.1$ at late times, and then to vanish again in the infinite future; at variance with the standard cosmology, and thanks to the exponential term related to the noise, $\Omega_k$ is allowed to stay relatively small even if $1-\Omega_m$ increases toward late times, thus never dominating the cosmic energy budget (see Figure 4 in \cite{Lapi2023}). However, in this interpretation the definition of distances in the ensemble-averaged cosmology is somewhat problematic, since the curvature component features a non-standard evolution and its relationship with an overall ensemble-averaged metric is non-trivial. In any case, the above discussion is reminiscent of deterministic backreaction models in a general relativistic setting, where the concepts of an emergent dark energy and an emergent curvature are often identified, as pioneered by \cite{Buchert2000b,Buchert2008} and numerically demonstrated by \cite{Bolejko2018}.

We also note that the $\eta$CDM model correctly reproduces as limiting cases the expected behavior in absence of the noise ($\zeta\approx 0$) or in a Milne's Universe devoid of matter/radiation components ($\Omega_{m,\gamma}\approx 0$). In both cases the exponential in Equation (\ref{eq|constr}) goes to unity, so that the emergent dark energy is just standard curvature ($\Omega_\eta=\Omega_k$) and the evolution closely resembling an Einstein de Sitter (with $\Omega_m\approx 1$ and $\Omega_\kappa\approx 0$) or a purely (anti-)de Sitter Universe (with $\Omega_m\approx 0$ and $\Omega_\kappa\approx 1$) is recovered. 

We have still to better justify the name `dark energy' for the component $\Omega_\eta$ defined above by deriving its equation of state. To this purpose it is instructive to rewrite all the original equations in terms of the triple $(h,\Omega_m,\Omega_\eta)$ by eliminating $\Omega_\gamma=1-\Omega_m-\Omega_\eta$. After some algebraic manipulations, we get
\begin{equation}\label{eq|etaCDM}
\left\{
\begin{aligned}
\dot{h} &= h^2\,\left(-2+\cfrac{\Omega_m}{2}+\Omega_\eta\right)+\cfrac{\zeta^2}{2}\,(1-\Omega_\eta)\, h^{2\alpha+1}\, \left[1-\cfrac{1-\alpha}{2}\, (1-\Omega_\eta)\right]\\
\dot{\Omega}_m &=\Omega_m\, h\, (1-\Omega_m-2\,\Omega_\eta)+\zeta^2\, \Omega_m\, \Omega_\eta\, h^{2\alpha}\, \left[1-\cfrac{4-\alpha}{2}\, (1-\Omega_\eta)\right]\\
\dot{\Omega}_\eta &=\Omega_\eta\, h\, (2-\Omega_m-2\,\Omega_\eta)-\zeta^2\,\Omega_\eta\,(1-\Omega_\eta)\, h^{2\alpha}\, \left[1-\cfrac{4-\alpha}{2}\, (1-\Omega_\eta)\right]~.
\end{aligned}
\right.
\end{equation}
We notice that at late times when structure formation develops, $\Omega_\eta\simeq 1-\Omega_m$ and $\dot \Omega_\eta\simeq -\dot\Omega_m$ apply, consistently with the above Equations. Now using $\dot\rho/\rho = (\dot\Omega/\Omega) + 2\, (\dot h/h) = -3\, h\,(1+w)$ we obtain:
\begin{equation}\label{eq|EoS}
\left\{
\begin{aligned}
w_m &= -\frac{\zeta^2}{3}\,h^{2\alpha-1}\,\left[1-\frac{1-\Omega_\eta}{2}\,(3\Omega_\eta+1-\alpha)\right]\simeq
-\frac{\zeta^2}{6}\,h^{2\alpha-1}\,\left[2-(4-\alpha)\, \Omega_m+3\,\Omega_m^2)\right]~,\\
&\\
w_\eta &=-\frac{1}{3}-\frac{\zeta^2}{2}\,h^{2\alpha-1}\,(1-\Omega_\eta)^2\simeq -\frac{1}{3}-\frac{\zeta^2}{2}\,h^{2\alpha-1}\,\Omega_m^2~,
\end{aligned}
\right.
\end{equation}
with the second expressions holding at late times when a non-zero $\Omega_\eta\simeq 1-\Omega_m$ applies. 
The first point to notice is that $\Omega_\eta$ features an equation of state $w_\eta\leq -1/3$ at all times, and so always tends to induce an accelerating behavior, justifying its name as emergent dark energy. As described earlier $w_\eta\to -1/3$ holds at early times to resemble a small curvature-like component before the noise term kicks in to enforce  acceleration.  

Moreover, the noise also alters the matter equation of state, which starts with the usual $w_m\approx 0$ at early times (i.e., pressureless dust), but progressively lowers toward $w_m\lesssim -1/3$ at late times. We stress that the modification of the matter equation of state is just apparent: the $\eta$CDM model relies on the standard cold DM scenario, and the intrinsic properties of the particles will plainly not be affected by the stochasticity. The matter equation of state changes since one insists to describe the ensemble-averaged energy density evolution in terms of a standard mass-energy conservation equation. The negative $w_m$ here just means that the dilution due to cosmic expansion for the ensemble-averaged matter density becomes slower than the standard trend. 

This often occurs in cosmological models with strong coupling in the dark sector (e.g., \cite{Mainini2007, DiValentino2020, Aich2023,vanderWesthuizen:2023hcl, Gleyzes:2015pma, Benisty:2018qed, Anagnostopoulos:2019myt,Giare:2024ytc}): in fact, the noise term in the last two Equations (\ref{eq|etaCDM}) can be viewed as describing a very specific form of dark coupling. In particular, such a noise-driven interaction makes an equal and opposite contribution to the evolution of $\Omega_m$ and $\Omega_\eta$, and renders a kind of energy transfer from DM to the emergent dark energy or viceversa, depending on its sign. At sufficiently high redshift when $\Omega_m\gtrsim 2/(4-\alpha)$ there is a net energy transfer from DM to dark energy (related to voids expanding fast and rapidly tending to dominate the Universe in volume), that actually reinforces the lowering of $\Omega_m$ and the rise of $\Omega_\eta$ with respect to the case without noise. At later times instead when $\Omega_{m}\lesssim 2/(4-\alpha)$ the situation is reversed (matter clustering proceeds along the structure of the cosmic web while void expansion progressively stalls since almost all the volume has been occupied) so that the evolution of the two components $\Omega_m$ and $\Omega_\eta$ considerably slows down, to the point of asymptotically saturating in the future (see also Section \ref{sec|coincidence}).

\subsection{\texorpdfstring{$\eta$CDM}{etaCDM} cosmography}\label{sec|cosmography}

The cosmological evolution at late times in the $\eta$CDM model can be better discerned by inspecting the cosmographic parameters. The deceleration parameter $q$ is routinely defined as 
\begin{equation}
q \equiv -\frac{\ddot{a}\,a}{\dot a^2} = -1-\frac{\dot h}{h^2}\;. 
\end{equation}
At late times, its expression in the $\eta$CDM model is explicitly given by 
\begin{equation}\label{eq|etacdm_q0}
q_{\eta\rm CDM}=\frac{\Omega_m}{2}-\frac{\zeta^2}{2}\, \Omega_m\, h^{2\alpha-1}\, \left(1-\frac{1-\alpha}{2}\, \Omega_m\right)~;
\end{equation}
for $\alpha\gtrsim -1$, $\zeta\sim 1$ of order unity and reasonable values of $h_0$ and $\Omega_{m,0}$, an accelerated expansion with the observed $q_0\lesssim -0.5$ can be easily obtained. For reference, in $\Lambda$CDM $q_{\Lambda\rm CDM}=-1+3\,\Omega_m/2$.

At higher-order, the jerk is defined as
\begin{equation}
j \equiv \frac{\dddot{a}\,a^2}{\dot a^3} = 1+3\,\frac{\dot h}{h^2}+\frac{\ddot h}{h^3} = q+ 2\,q^2-\frac{\dot q}{h}
\end{equation}
and its expression in $\eta$CDM reads
\begin{equation}\label{eq|etacdm_jerk}
\begin{aligned}
j_{\eta\rm CDM} & = \Omega_m \left\{1-\frac{\zeta^2}{8}\, h^{2\, \alpha-1} \left[8 \,(1+\alpha)+2 \,\left(-7+3 \alpha+2 \alpha^2\right)\, \Omega_m+\left(9-5\alpha+2\alpha^2\right)\, \Omega_m^2\right]\right.+\\
&\left.+ \frac{\zeta^4}{16}\, h^{4 \alpha-2}\, \left[8+4\, (-7+5 \alpha)\, \Omega_m+12\, \left(3-3 \alpha+\alpha^2\right) \Omega_m^2+\left(-15+20\alpha-7\alpha^2+2 \alpha^3\right)\, \Omega_m^3\right]\right\}\; ;
\end{aligned}
\end{equation}
for reference, in $\Lambda$CDM $j_{\Lambda\rm CDM}=1$.

In addition, the snap is defined as
\begin{equation}
s \equiv \frac{\ddddot{a}\,a^3}{\dot a^4} = 1+6\,\frac{\dot h}{h^2}+4 \frac{\ddot h}{h^3}+\frac{3 \dot{h}^2+\dddot{h}}{h^4} = -2j-3q\,j+\frac{\dot j}{h}
\end{equation}
and its expression in $\eta$CDM reads
\begin{equation}\label{eq|etacdm_snap}
\begin{aligned}
s_{\eta\rm CDM} & = \Omega_m\, \Biggl\{-3-\frac{\Omega_m}{2} +\frac{\zeta^2}{8}\, h^{2\alpha-1}\, \Biggl[(24+32\alpha+16\alpha^2)+2\,(-27+\alpha+16\alpha^2+4\alpha^3)\, \Omega_m +\\
&+(60-20\alpha+8\alpha^3)\, \Omega_m^2+(-18+19\alpha-9\alpha^2+2\alpha^3)\,\Omega_m^3\Biggr]-\frac{\zeta^4}{32}\, h^{4\alpha-2}\, \Biggl[(48 + 96\alpha)+\\
&+(-296-80\alpha+240\alpha^2)\,\Omega_m+ 8\,(92-45\alpha-18\alpha^2+18\alpha^3)\,\Omega_m^2+(-784+754\alpha+\\
&-210\alpha^2+24\alpha^3+24\alpha^4)\,\Omega_m^3+(303-364\alpha+221\alpha^2-64\alpha^3+12\alpha^4)\,\Omega_m^4\Biggr]+\\
&+\frac{\zeta^6}{64}\, h^{6 \alpha-3}\, \Biggl[32+(-304+240\alpha)\,\Omega_m+16\, (69 -78\alpha+ 26\alpha^2)\,\Omega_m^2 + 4\, (-471 + \\
&+636 \alpha - 319 \alpha^2 + 70 \alpha^3)\,\Omega_m^3+
2\,(759 - 1156 \alpha + 693 \alpha^2 - 228 \alpha^3 + 40 \alpha^4)\,\Omega_m^4+\\
&+(-465+785\alpha-497\alpha^2+219\alpha^3-50\alpha^4+8\alpha^5)\, \Omega_m^5\Biggr] \Biggr\}\; ;
\end{aligned}
\end{equation}
for reference, in $\Lambda$CDM $s_{\Lambda\rm CDM}=1-9\,\Omega_m/2$. 

If one inserts in Equations (\ref{eq|etacdm_q0}-\ref{eq|etacdm_jerk}-\ref{eq|etacdm_snap}) the matter density parameter $\Omega_m(z)$ and the Hubble rate $h(z)$ as a function of redshift obtained by solving the system of Equations (\ref{eq|etaCDM}), then the deceleration, jerk and snap parameters at any epochs are directly obtained (only requirements is that radiation energy density is small, and we have checked that this implies a range of validity $z\lesssim 30$). On the other hand, from the same Equations above one can compute  $q_0$, $j_0$ and $s_0$ at $z\approx 0$ as a function of the present matter density $\Omega_{m,0}$, Hubble constant $h_0$ and of the noise parameters $\zeta$ and $\alpha$. 

These can in turn be exploited to obtain cosmographic approximations at low redshift $z\lesssim 0.5$ for the Hubble rate and the luminosity distance\footnote{Since the inhomogeneities/anisotropies described by the noise term are expected to be small, it is a reasonable approximation to assume that the ensemble-averaged Universe is described by a Friedman-Robertson-Walker metric, and to adopt the standard redshift-distance relation \cite{Koksbang2019,Koksbang2019b}.} (e.g., \cite{Visser2005,Capoziello2011}). Specifically, one has:
\begin{equation}\label{eq|hz_cosmograp}
\begin{aligned}
h(z) &\simeq h_0\,\Biggl[1+y\,(1+q_0)+\frac{y^2}{2}\,(2+j_0+2 q_0-q_0^2)+\frac{y^3}{6}\,(6+6 q_0-3 q_0^2+3 q_0^3+ 3 j_0 -4 j_0 q_0-s_0)\Biggr] \; ,
\end{aligned}
\end{equation}
and
\begin{equation}\label{eq|DL_cosmograp}
d_L(z) \simeq \frac{c}{H_0}\,y\, \Biggl[1+\frac{3\,y}{2}\,(1-q_0)+\frac{y^2}{6}\,(11-5 q_0+3 q_0^2-j_0)+\frac{y^3}{24}\,(50-26 q_0+21 q_0^2-15 q_0^3-7 j_0+10 j_0 q_0+s_0)\Biggr] \; ,
\end{equation}
where $y\equiv z/(1+z)$. From the latter expression the proper distance $d_M=d_L/(1+z)$ and the angular diameter distance $d_A=d_L/(1+z)^2$ can be easily derived. These analytic expressions can be useful for the future analysis of cosmological datasets.

\subsection{Fate of the Universe and coincidence problem}\label{sec|coincidence}

Remarkably, in the infinite future the ensemble-averaged evolution in $\eta$CDM features an attractor solution, that can be found by setting $\dot h=\dot\Omega_m=\dot\Omega_\eta\approx 0$ in Equations (\ref{eq|etaCDM}), to yield
\begin{equation}\label{eq|attsol}
\left\{
\begin{aligned}
\Omega_{m,\infty} &= \frac{\alpha-4+\sqrt{(\alpha-4)^2+12}}{3}~,\\
&\\
\Omega_{\eta,\infty} &= 1-\Omega_{m,\infty}~,\\
&\\
h_\infty &= \left[\zeta^2\,\left(1-\frac{4-\alpha}{2}\, \Omega_{m,\infty}\right)\right]^{\frac{1}{1-2\alpha}}~;
\end{aligned}
\right.
\end{equation}
in addition one can show from Equations (\ref{eq|EoS}) that $w_m\approx w_\eta\approx -1$, $q_\infty\approx -1$, $j_\infty\approx 1$ and $s_\infty\approx 1$. To some extent, the noise will originate a kind of inflationary behavior in the infinite future, with all the relevant components behaving like a cosmological constant, i.e. featuring a constant energy density. Note that the condition $\Omega_m < 2/(4-\alpha)$ must hold at late cosmic times for the third expression above to be meaningful, and for this attractor solution to set in. As already mentioned in Section \ref{sec|emergentDE}, the presence of this attractor solution at late times can be traced back to a kind of equilibrium in the energy exchange between $\Omega_m$ and $\Omega_\eta$, or in more physical terms between the clustering of matter along the structures of the cosmic web and the increasing dominance in volume by the voids.

As also pointed out in \cite{Lapi2023}, it is easy to check that for $\alpha\gtrsim -1$ and $\zeta\sim 1$ the cosmological parameters will hover around values similar to the current ones for an infinite amount of time in the future. This is strongly at variance with many other cosmological models, where the matter and dark energy densities have an opposite trend with cosmic time (e.g., in $\Lambda$CDM one has $\Omega_m\approx 1$ and $\Omega_\Lambda\approx 0$ in the remote past while $\Omega_m\approx 0$ and $\Omega_\Lambda\approx 1$ in the infinite future), thus 
bringing about the cosmic coincidence issue of why nowadays these quantities have similar values. In $\eta$CDM, given that the present values will stay put for an infinite amount of time in the future, the cosmic coincidence is solved without recurring to anthropic considerations, apart from the trivial fact that a human being must be here and now to raise such an issue. 

\begin{figure}[tbp]
    \centering
    \includegraphics[scale=0.45]{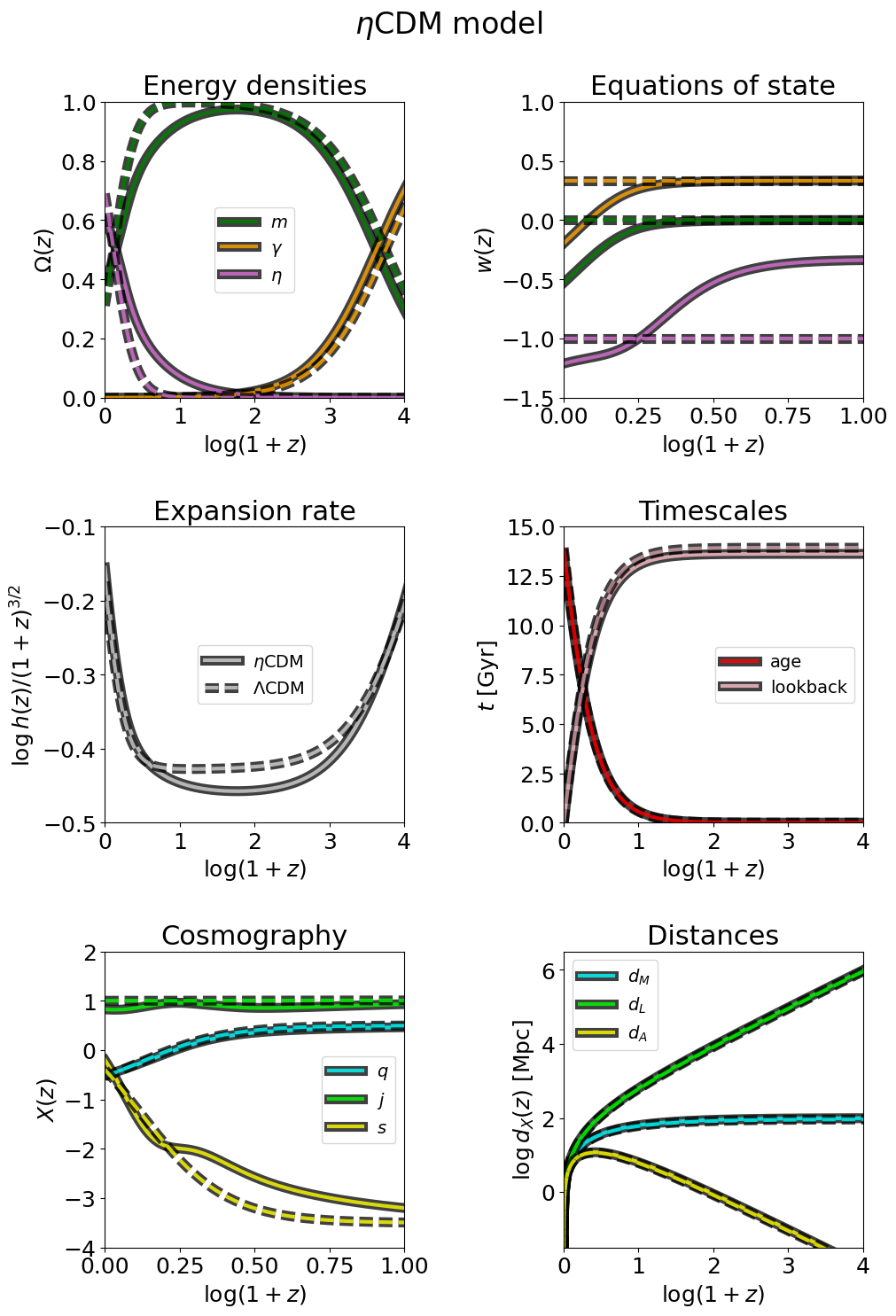}
    \caption{Ensemble-averaged evolution of the Universe in the $\eta$CDM model (solid lines). The different panels illustrate the energy densities (top left), equations of state (top right), expansion rate (middle left), timescales (middle right), cosmographic quantities (bottom left) and cosmological distances (bottom right). In the top panels, green lines refer to matter, orange lines to radiation and  magenta lines to the (emergent) dark energy; in the middle right panel pink line shows the lookback time and red line shows the age of the Universe; in the bottom left panel, cyan lines show the deceleration parameter $q$, green lines show the jerk parameter $j$ and yellow lines the snap parameter $s$; in the bottom right panel, cyan lines show the transverse comoving distance $d_M$, green line the luminosity distance $d_L$ and yellow lines the angular distance $d_A$. In all panels the standard $\Lambda$CDM model (dashed lines) is also reported for reference.}
    \label{fig|evo}
\end{figure}

\subsection{\texorpdfstring{$\eta$CDM}{etaCDM} vs. \texorpdfstring{$\Lambda$CDM}{etaCDM} evolution}\label{sec|evo}

In Figure \ref{fig|evo} we illustrate the evolution with redshift of various relevant quantities in the $\eta$CDM model, and compare them with the analogous ones in $\Lambda$CDM. For $\eta$CDM these outcomes are obtained just by integrating Equations (\ref{eq|etaCDM}) backward in time from given initial conditions at the present epoch. Specifically, we choose the set of cosmological parameters $h_0\approx 0.7$, $\Omega_{m,0}\approx 0.4$, $\zeta\approx 1.5$, $\alpha\approx -0.75$ that will turn out  to fit quite well cosmological observables (see Section \ref{sec|results}). For the sake of definiteness, in $\Lambda$CDM we adopt the \texttt{Planck} bestfits \cite{Planck2020}. 

It is seen that the expansion rate, the energy densities, timescales, distances are very similar in the two models, with the main deviations occurring at late times when the noise becomes important. Perhaps the most striking difference can be identified in the evolution of the equation of state for the dark energy component, which is varying with redshift in $\eta$CDM, while it is constant $w_\Lambda\approx -1$ in $\Lambda$CDM (at least in its basic version). This is an occurrence that could be tested by the \texttt{Euclid} mission via weak lensing and galaxy clustering analysis. We also highlight the appreciably different evolutions of the cosmographic parameters at intermediate to high $z$, especially the jerk and the snap. These could provide independent tests to be performed via searches for transient objects (e.g., supernovae) with future surveys like those planned on the \texttt{Nancy Grace Roman Space Telescope}. Another potentially interesting diagnostics to differentiate the two models concerns the growth rate of cosmic structures vs. that of the background, and will be discussed in Section \ref{sec|cojoin}.

\section{Tuning noise and fitting data}\label{sec|data}

We will now turn to set the parameters of the $\eta$CDM model and to investigate its overall performance in reproducing a wealth of recent cosmological observations. 

\subsection{Data and Analysis} \label{sec|datasets}
We focus on the following datasets:

\begin{itemize}

\item \emph{Baryon acoustic oscillations (BAO)}. We exploit the \texttt{DESI} BAO sample of $7$ measurements in the redshift range $z\sim 0.1-4$ by \cite{Adame2025} to fit for the ratios $d_M/r_d$, $d_H/r_d$ and $d_V/r_d$ between the transverse comoving distance 
\begin{equation}
d_M(z)=c\,\int_0^z\cfrac{{\rm d}z'}{H(z')}~,
\end{equation}
the Hubble distance $d_H(z)=c/H(z)$ and the angle-averaged distance $d_V(z)=[z\,d_M^2\,d_H]^{1/3}$ to the scale of the drag epoch sound horizon $r_d$.

\item \emph{Type-I$a$ supernovae (SN)}. We exploit the \texttt{Pantheon+} sample of $\gtrsim 1500$ unique type-I$a$ SN in the redshift range $z\sim 0.01-2.3$ by \cite{Scolnic2022} to fit for the magnitude-redshift relation $m_b = 5\, \log[d_L(z)/\mathrm{Mpc}]+25+M_b$, where $d_L(z)=d_M\,(1+z)$ is the luminosity distance and $M_b$ is the fiducial absolute magnitude of a type-I$a$ SN calibrated via the distance ladder (e.g., using Cepheids as an anchor). 

\item \emph{Structure growth rate (SGR).} We exploit the sample of $18$ robust and independent\footnote{Correlations amongst the WiggleZ \cite{Blake2012} data are taken into account, although they do not have a substantial impact in the final inference.} measurements from different galaxy surveys in the redshift range $z\sim 0.02-1.4$ (see \cite{Nesseris2017}) to fit for the 
structure growth rate $f\sigma_8(z)=\sigma_8\,\left|{\rm d}\delta_+/{\rm d}\ln a\right|$, where $\sigma_8$ is the rms amplitude of the perturbation spectrum on a scale of $8\, h^{-1}$ Mpc and $\delta_+$ is the growing mode of the linear density contrast for the matter perturbations. The evolution of the latter quantity is computed by solving the differential equation (see Appendix \ref{app|linpert})
\begin{equation}\label{eq|linpert}
\delta''+(2+H'/H)\,\delta'-(3/2)\,\Omega_m\,\delta=0   
\end{equation}
where the prime denotes differentiation with respect to $\ln a$;
the adopted initial conditions are set at $\ln a_{\rm ini}\approx -3$ where $\delta_{\rm ini}=\delta_{\rm ini}'=a_{\rm ini}$ since we expect that at sufficiently high redshift (yet in the matter dominated epoch) the noise is negligible and thus $\delta=a$ applies as in the standard model. 

\item \emph{Cosmic chronometers (CC)}. We exploit the sample 
of $33$ measurements by \cite{Moresco2024} for the redshift-dependent Hubble rate $H(z)$ as determined from differential ages of early-type galaxies. We use the full covariance
matrix, taking into account modeling uncertainties, mainly related to the choice of the initial mass function, of stellar libraries and stellar population synthesis models (see \cite{Moresco2022} for details).

\end{itemize}

For the analysis we adopt a Bayesian framework, characterized by the parameter set $\theta\equiv \{h_0,\Omega_{m,0},\zeta,\alpha,r_d,M_b,\sigma_8\}$. We implement a Gaussian log-likelihood
\begin{equation}\label{eq|likelihood}
\ln \mathcal{L}(\theta) = -\chi^2(\theta)/2~,
\end{equation}
where the chi-square $\chi^2(\theta)=\sum_{ij}\,[\mathcal{M}(\theta,z_i)-\mathcal{D}(z_i)]\,\mathcal{C}_{ij}^{-1}\, [\mathcal{M}(\theta,z_j)-\mathcal{D}(z_j)]$ is obtained by comparing our model expectations $\mathcal{M}(\theta,z_i)$ to the data values $\mathcal{D}(z_i)$, summing over different observables at their respective redshifts $z_i$ and taking into account the variance-covariance matrix $\mathcal{C}_{ij}$ among redshift bins. We adopt flat priors $\pi(\theta)$ on the parameters $h_0\in [0.5,1.0]$, $\Omega_{m,0}\in [0,1]$, $\zeta\in  [0,3]$ and $\alpha\in [-2,0]$, $r_d\, [\rm{Mpc}]\in [120,160]$, $M_b\, [\rm{mag}]\in [-20.5,-18.5]$, $\sigma_8\in [0.6,1.2]$. The bound $\Omega_{m,0}<2/(4-\alpha)$ is also set to ensure late-time physical solutions solving the cosmic coincidence problem (see Section \ref{sec|coincidence}). 

Moreover, when specified explicitly we add the following robustly accepted priors:

\begin{itemize}

\item[$\square$] \emph{Horizon scale at drag epoch $r_d$}. We impose a prior on the horizon scale at the drag epoch. Since in $\eta$CDM the pre-recombination physics is unaltered with respect to the standard cosmology (noise is negligible at high redshift) we rely on the $\Lambda$CDM value from \cite{Planck2020} and adopt a Gaussian prior on $r_d$ [Mpc] with mean $147.18$ and dispersion $0.29$.

\item[$\square$] \emph{Type-I$a$ SN zero-point magnitude}. We impose a prior\footnote{Note that the prior on $M_b$ is correlated with the $z \lesssim 0.01$ SNe in the \texttt{Pantheon+} dataset, which are thus excluded from the SN likelihood.} on the absolute magnitude $M_b$ of a type-I$a$ SN derived from the distance ladder via Cepheids, variable asymptotic giant branch stars and the tip of the red giant branch. We adopt a Gaussian prior on $M_b$ with mean $-19.28$ and dispersion $0.03$, as suggested by \cite{Riess2024} taking into account classic \texttt{HST} measurements and recent \texttt{JWST} cross-validation.

\item[$\square$] \emph{CMB first-peak angular scale}. We impose a prior on the angular scale of the first peak in the CMB temperature spectrum $\theta_\star\approx r_\star/d_M(z_\star)$, where $r_\star$ is the comoving sound horizon at recombination and $d_M(z_\star)$ is the transverse comoving distance at the recombination redshift $z_\star$. As above we assume a $\Lambda$CDM pre-recombination physics and thus set $r_\star\approx r_d/1.018$\footnote{{The ratio $r_d/r_\star = 1.01841 \pm 0.00033$ of the sound horizon at last scattering ($r_\star$) and at the drag epoch ($r_d$) is completely correlated with the baryon density $\Omega_{b}\,h^2 = 0.02237 \pm 0.00015$ \cite{Planck2020}. By adopting a fixed value for $r_d/r_\star$ implies we are setting the baryon contribution to the total matter density, which is the only free parameter. Note that we also neglect the overall correlation between $\theta_\star$ and $r_d$ when imposing both priors; actually the mild correlation between these parameters (with a coefficient $\sim 0.17$) has no major impact on our final inference outcomes.}} and $z_\star\approx 1090$ from \cite{Planck2020}. Then we follow \cite{Adame2025} and adopt a Gaussian prior on $100\,\theta_\star$ with mean $1.0411$ and dispersion $0.0005$, which is approximately twice the \texttt{Planck} \cite{Planck2020} uncertainty.

\item[$\square$] \emph{Age of the Universe}.
We impose a prior on the age of the Universe from globular cluster dating. We adopt a Gaussian prior on $t_0$ [Gyr] with mean $13.5$ and dispersion $0.4$ as suggested from \cite{Valcin2021}.

\end{itemize}

We sample the parameter posterior distributions $\mathcal{P}(\theta) \propto \mathcal{L}(\theta)\,\pi(\theta)$ via the MCMC Python package \texttt{emcee} \cite{emcee}, running it with $10^5$ steps and $100$ walkers; each walker is initialized with a random position extracted from the priors discussed above. To speed up convergence, we adopt a mixture of differential evolution and snooker moves of the walkers, in proportion of $0.8$ and $0.2$ respectively, that emulates a parallel tempering algorithm. After checking the auto-correlation time, we remove the first $30\%$ of the flattened chain to ensure burn-in.

\begin{figure}[t!]
    \centering
    \includegraphics[width=\textwidth]{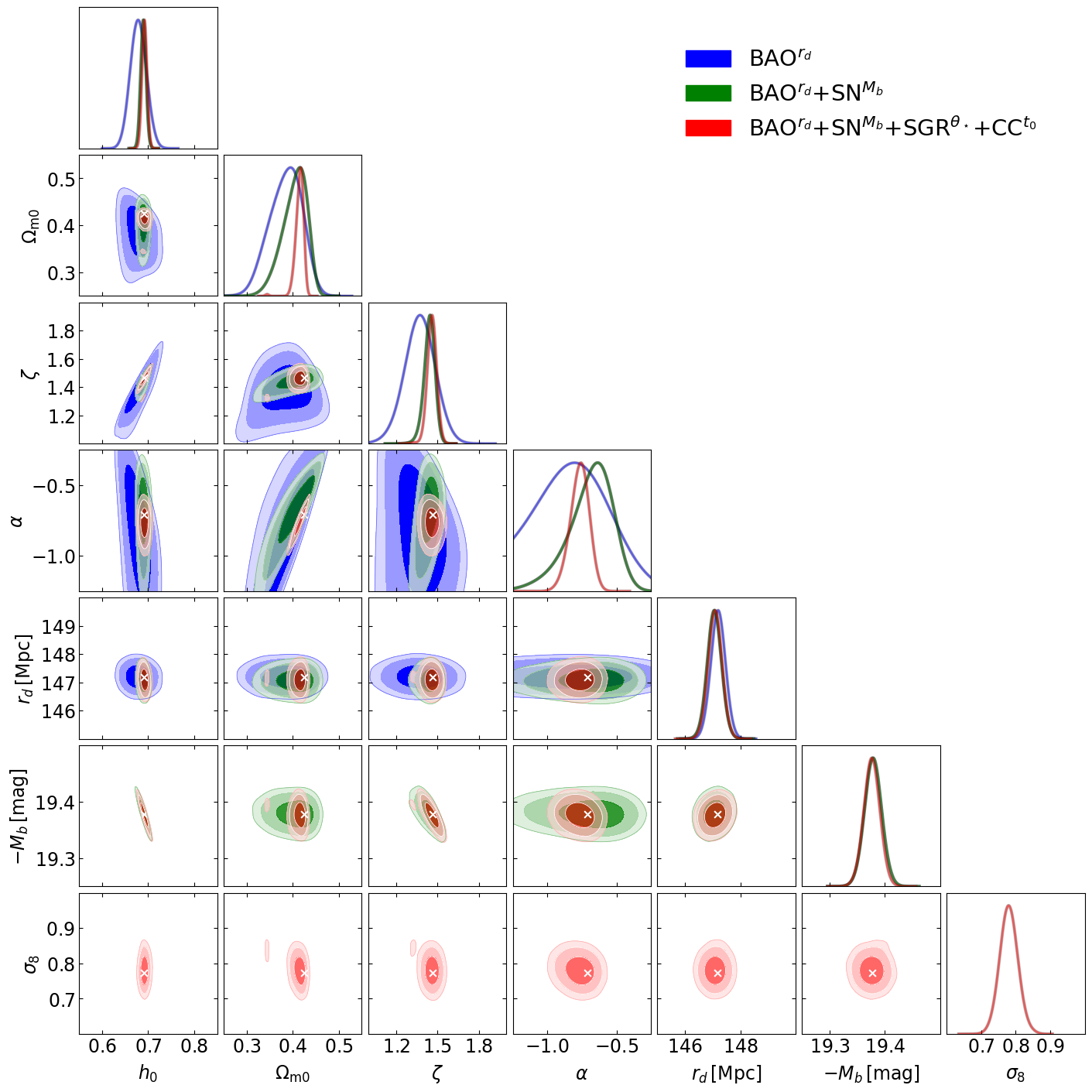}
    \caption{MCMC posterior distributions of the (normalized) Hubble constant $h_0$, present matter density parameter $\Omega_{m,0}$, noise parameters $\zeta$ and $\alpha$, horizon at the drag epoch $r_d$, type-I$a$ SN zero-point magnitude $M_b$ and amplitude of the linear matter power spectrum $\sigma_8$. Coloured contours/lines refer to the analysis with different dataset(s): blue is for BAO with the prior on $r_d$; green adds SN with the prior on $M_b$; red adds SGR and CC with the priors on $\theta_\star$ and $t_0$. The marginalized distributions are in arbitrary units (normalized to 1 at their maximum value). The white crosses highlights the position of the best fit values for the full dataset.}
    \label{fig|MCMC}
\end{figure}

{\renewcommand{\arraystretch}{1.75}  \setlength{\tabcolsep}{2.5pt}
\begin{table}[htbp]
\footnotesize
\centering
\begin{tabular}{lcccccccc}
\hline
\hline
Dataset(s) & $h_0$ & $\Omega_{m,0}$ & $\zeta$ & $\alpha$ & $r_d$ [Mpc] & $-M_b$ [mag] & $\sigma_8$\\
\hline
BAO$^{r_d}$ & $0.68^{+0.02}_{-0.02}$ & $0.39^{+0.04}_{-0.03}$ & $1.37^{+0.11}_{-0.11}$ & $-0.89^{+0.37}_{-0.26}$ & $147.2^{+0.3}_{-0.3}$ & $-$ & $-$\\

BAO$^{r_d}+$SN$^{M_b}$ & $0.69^{+0.01}_{-0.01}$ & $0.40^{+0.03}_{-0.02}$ & $1.44^{+0.04}_{-0.04}$ & $-0.69^{+0.19}_{-0.12}$ & $147.1^{+0.3}_{-0.3}$ & $19.38^{+0.02}_{-0.02}$ & $-$\\

BAO$^{r_d}+$SN$^{M_b}+$SGR$^{\theta_\star}$+CC$^{t_0}$ & $0.69^{+0.01}_{-0.01}$ & $0.41^{+0.01}_{-0.01}$ & $1.46^{+0.04}_{-0.03}$ & $-0.77^{+0.08}_{-0.06}$ & $147.1^{+0.3}_{-0.3}$ & $19.38^{+0.02}_{-0.02}$ &$0.78^{+0.03}_{-0.03}$ \\

\hline

SN$^{M_b}$ & $0.72^{+0.01}_{-0.01}$ & $0.26^{+0.09}_{-0.09}$ & $1.60^{+0.15}_{-0.23}$ & $-0.90^{+0.77}_{-0.39}$ & $-$ & $19.28^{+0.03}_{-0.03}$ & $-$\\

BAO$+$SN$+$SGR$^{\theta_\star}$+CC$^{t_0}$ & $0.68^{+0.02}_{-0.02}$ & $0.40^{+0.03}_{-0.01}$ & $1.36^{+0.07}_{-0.07}$ & $-0.87^{+0.15}_{-0.10}$ & $147.6^{+2.6}_{-2.6}$ & $19.40^{+0.04}_{-0.04}$ & $0.78^{+0.05}_{-0.05}$ \\
\hline
\hline
\end{tabular}
\caption{Marginalized posterior estimates (mean and $1\sigma$ confidence intervals are reported) from the MCMC analysis of different cosmological datasets with the $\eta$CDM model. Columns report the values of the (normalized) Hubble constant $h_0$, present matter density $\Omega_{m,0}$, noise parameters $\zeta$ and $\alpha$, horizon scale at drag epoch $r_d$, SN I$a$ absolute magnitude $M_b$, and amplitude of linear matter perturbation spectrum $\sigma_8$. The rows refer to the different dataset combinations exploited for the fit: first line is for BAO with the prior on $r_d$; second line adds SN with the prior on $M_b$; third line adds SGR and CC with the priors on $\theta_\star$ and $t_0$; forth line is for only SN with the prior on $M_b$, and last line is for the combined datasets (as in the third line) but with no priors on $r_d$ and $M_b$.}
\label{tab|results} 
\end{table}}

\begin{figure}[t!]
    \centering
    \includegraphics[width=\textwidth]{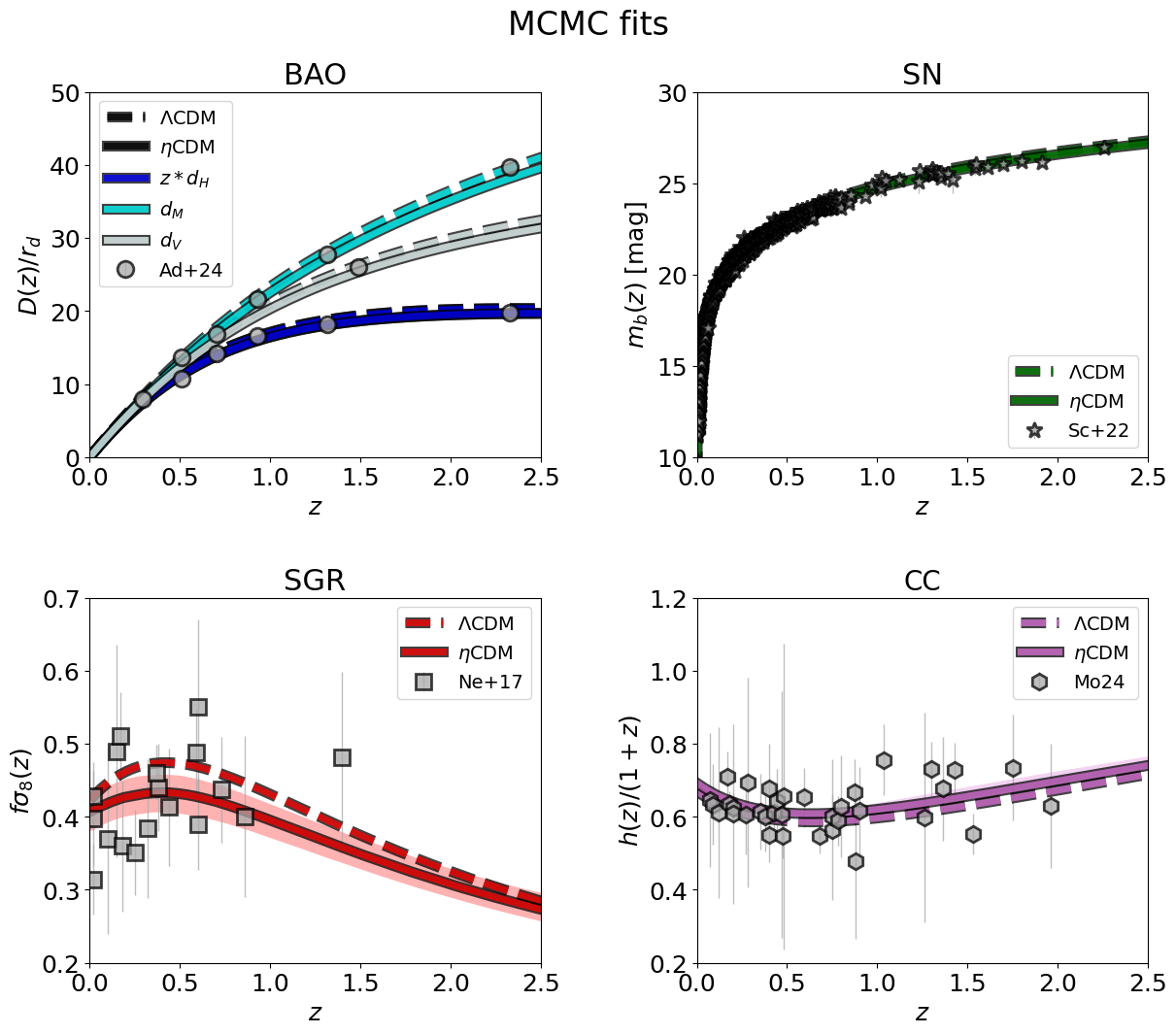}
    \caption{MCMC fits to the BAO distance ratios (top left panel) $d_H(z)/r_d$ (blue) $d_M(z)/r_d$ (cyan) $d_V(z)/r_d$ (azure), type-I$a$ SN magnitude vs. redshift $m_b(z)$ (top right), structure growth rate $f\sigma_8(z)$, cosmic chronometer determination of the (normalized) Hubble rate $h(z)$ in the $\eta$CDM model. Colored lines illustrate the median, and the shaded areas show the $2\sigma$ credible interval from sampling the posterior distributions. For reference, the dashed lines display the standard $\Lambda$CDM model with parameter set from \texttt{Planck} \cite{Planck2020}. Data are from \cite{Adame2025} (circles), \cite{Scolnic2022} (stars), \cite{Nesseris2017} (squares), and \cite{Moresco2024} (hexagons).}
    \label{fig|fits}
\end{figure}

\subsection{Results}\label{sec|results}   

The results of our Bayesian analysis are displayed in Figure \ref{fig|MCMC} and reported in Table \ref{tab|results}. Specifically, in Figure \ref{fig|MCMC} we illustrate the MCMC posterior distributions on the fitting parameters for various datasets (color-coded); the marginalized distributions are normalized to unity at their maximum, and the crosses mark the best-fit values of the parameters for the joint analysis. In Table \ref{tab|results} we summarize the marginalized posterior estimates of the parameters (median values and $1\sigma$ credible intervals).

We first analysed separately BAO data with the prior on $r_d$ from CMB (first line in Table \ref{tab|results}) and SN data with the prior on $M_b$ from the cosmic ladder (fourth line in Table \ref{tab|results}). From BAO we find rather well constrained values of $h_0\approx 0.68$ and $\Omega_m\approx 0.4$ and noise strength $\zeta\approx 1.4$, while the noise evolution parameter $\alpha\approx -0.9$ is loosely estimated. From SN we find a Hubble constant $h_0\approx 0.72$ (which reflects the prior on $M_b$) higher than BAO and a matter density $\Omega_{m,0}\approx 0.3$ lower than BAO though not particularly well constrained, a noise strength $\zeta\approx 1.6$ slightly larger than BAO, and no robust inference on the noise evolution parameter $\alpha$. However, all the parameters (and in particular $h_0$) from the BAO and SN analyses are consistent within $\sim 1.5\sigma$, so that no statistically-significant Hubble tension exists in $\eta$CDM. 

Provided that, we combine the BAO and SN datasets to improve our constraining power, especially in relation to the noise parameters (second line in Table \ref{tab|results}). In fact, both the noise strength $\zeta\approx 1.44$ and the noise evolution $\alpha\approx -0.7$ are considerably better constrained, though the latter retains a quite appreciable uncertainty. The Hubble constant moves to $h_0\approx 0.7$, with the related value of $r_d$ staying put with respect to the prior from CMB, while the type-I$a$ SN zero-point $M_b$ is shifted from the mean prior value of $\lesssim 2\sigma$.

Finally, we perform a joint analysis by adding the structure growth rate and cosmic chronometers data, with priors on the CMB $\theta_\star$ and on the age of the Universe $t_0$ from globular cluster dating (third line in Table \ref{tab|results}). 
This allows us to refine considerably the determination of the noise evolution parameter $\alpha\approx -0.8$, and to obtain a rather precise estimate for the normalization of the linear matter power spectrum $\sigma_8\approx 0.78$. The latter is found to be soundly consistent with the CMB expectation around $0.81$, so that no $f\sigma_8$ tension exists in the $\eta$CDM model. No strong degeneracies among the parameters emerge, apart from low-degree dependencies among $h_0$ vs. $\zeta$ vs. $M_b$, and between $\Omega_{m,0}$ vs. $\alpha$.

We repeat this joint analysis of the overall datasets without imposing priors on $r_d$ and $M_b$, to test whether these could bias significantly some of our inferences (last line in Table \ref{tab|results}). This is not the case, as all the parameters are found to be consistent with the priorized analysis within $\lesssim 1\sigma$; moreover, the uncertainties on all model parameters are marginally enlarged. Interestingly, the SN zero-point is very well constrained also in this case, while the $r_d$ determination has a substantially larger uncertainty. In addition, the value of $h_0$ which is now mainly guided by the CC data instead of the inverse-distance ladder (prior on $r_d$ is not assumed) is identical to the constraints imposed by the latter, indicating a high degree of consistency amongst datasets within the $\eta$CDM model.

Comparing with the previous joint analysis in \cite{Lapi2023}, the bestfit parameters found in this work, despite being consistent within $2\sigma$, show a few appreciable differences. The most noticeable ones were generally higher values $h_0\gtrsim 0.7$ and smaller values of $\alpha\lesssim -1$ found there. We have tested that such differences are driven by two reasons. The first is the much precise BAO datasets employed in the present analysis. Specifically, here we fit for the BAO distance scale ratios $d_M/r_d$, $d_H/r_d$ and $d_V/r_d$ while in \cite{Lapi2023} only spherically-averaged BAO data were considered. 
The second is the revised prior on $M_b$, that in turn strongly determines the value of and uncertainty on $h_0$ from SN. In \cite{Lapi2023} the extremely stringent prior from the Cepheid anchor by \cite{Riess2022} was employed, while here we adopt the more balanced one from \cite{Riess2024} that overall yields a lower $h_0$ with a slightly larger uncertainty. The combination of these two occurrences in \cite{Lapi2023} weighted low the BAO data and high the SN ones in the joint analysis, driving the bestfit toward a higher value of $h_0$ and a smaller (and more uncertain) value of $\alpha$. 

We stress the relevance of the SGR data considered in this work, which provide quite stronger constraint on $\alpha$ with respect to the BAO+SN combination alone. This is because $\alpha$ determines the redshift evolution of the noise, and in turn considerably affects the overall behavior of the Hubble parameter and of the growth factor in the redshift range probed by galaxy redshift surveys. In principle, further constraints 
on the noise parameters $\zeta$ and $\alpha$ can be provided by BAO-independent determinations of the Hubble rate such as those offered by cosmic chronometers. We include such data in our analysis, but their present uncertainties are so large that they marginally contribute to the overall parameter inference. The situation, hopefully, will improve in the near future with new datasets from galaxy surveys and a better control on systematics uncertainties related to stellar and galaxy evolution models in cosmic chronometer measurements (see \cite{Moresco2024} for an extended discussion).

In Figure \ref{fig|fits} the best fits (solid lines) and the $2\sigma$ credible intervals (shaded areas, in most cases barely visible because they are very narrow) sampled from the posteriors are projected onto the observables: BAO distance ratios, type-I$a$ SN magnitude vs. redshift diagram, structure growth rate $f\sigma_8(z)$, and Hubble rate $h(z)$ to be confronted with cosmic chronometers. The fits are always extremely good, with overall reduced $\chi^2_r\lesssim 1$. For reference, the outcomes on the same observables for the standard $\Lambda$CDM model with parameter set from \texttt{Planck} \cite{Planck2020} are also reported. All in all, the performances of the $\eta$CDM and $\Lambda$CDM model are very similar; this is also confirmed by Bayesian comparison testing in terms of the Bayesian evidence or of Bayes inference criterion, that as expected do not show any reasonable preference in favor of either models. Maybe the most relevance difference is in the evolution of the $f\sigma_8$ parameter with redshift, a potentially interesting probe that we will discuss more deeply in Section \ref{sec|cojoin}.

\begin{figure}[t!]
    \centering
    \includegraphics[width=0.85\textwidth]{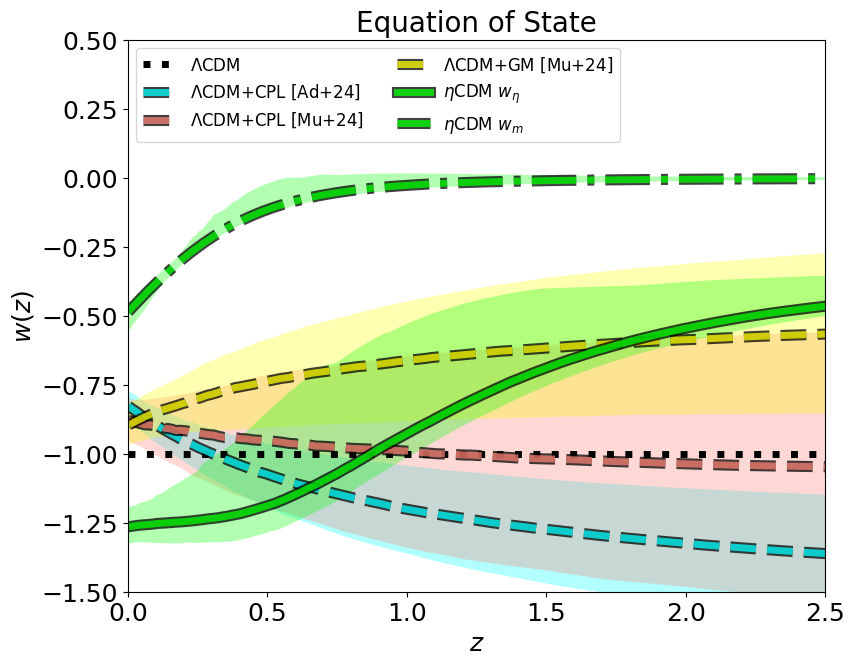}
    \caption{Evolution of the equation of state. Green is the result for the $\eta$CDM model from our analysis (solid for emergent dark energy $w_\eta$ and dot-dashed for matter $w_m$). Cyan is the result $w_\Lambda$ from the $\Lambda$CDM+CPL analysis of  BAO+SN data by \cite{Adame2025}, red from the $\Lambda$CDM+CPL analysis of BAO+SN+SGR data by \cite{Upala2024} and yellow from a $\Lambda$CDM+generalized CPL analysis of BAO+SN+SGR data by \cite{Upala2024}. For reference, dotted black line is the value $w_\Lambda=-1$ of the standard $\Lambda$CDM model.}
    \label{fig|EoS}
\end{figure}

\section{Discussion}\label{sec|discussion}

In this Section we focus on, and discuss some relevant consequences of the above analysis.

\subsection{Dark energy equation of state}\label{sec|DE}

In Figure \ref{fig|EoS} we show in green the equation of state for the emergent dark energy component in the $\eta$CDM model. Solid line is the median and shaded area is the $1\sigma$ confidence intervals from sampling the posterior distributions from our MCMC analysis of joint cosmological datasets. As anticipated already in Section \ref{sec|theory} the dark energy equation of state parameter $w_\eta$ has a decreasing trend from values close to $w_\eta\approx -1/3$ at high redshift down to values $w_\eta\lesssim -1$ at the present time (and then coming back asymptotically to $w_\eta\rightarrow -1$ in the infinite future). First of all, we stress that in the $\eta$CDM model one should not be strongly concerned by values of $w_\eta< -1$. In fact, while for a dynamical dark energy field these values would imply a violation of the null energy condition, in the $\eta$CDM model dark energy is just a description for the accelerated expansion of the Universe, emergent from the noise associated to structure formation.

The same is true for the matter component (reported as a green dot-dashed line in Figure \ref{fig|EoS}), whose equation of state is apparently modified from $w_m\approx 0$ at high redshift to values $w_m\lesssim -0.5$ at the present time (and toward $w_m\rightarrow -1$ in the infinite future). In fact, the properties of cold DM particles are clearly not affected by the noise associated to structure formation. What changes is the evolution of $\rho_m(z) = \Omega_m\, h^2$ with redshift, in such a way of originating an apparent decrease in $w(z)$ if one insists in describing the ensemble-averaged matter density with the standard fluid mass-energy conservation equation\footnote{Basically one is imposing the separate conservation of the stress energy tensor for each fluid component, while this is not the case at late times since matter and emergent dark energy are coupled. Plainly conservation of DM particles' number is not affected.}. 
As anticipated, such a behavior is often present in cosmological models with a strong-coupling in the dark sector (e.g., \cite{Mainini2007,Marra2016,DiValentino2020,Aich2023}), to which the ensemble-averaged $\eta$CDM evolution is similar, because dark energy is emergent from structure formation hence it depends on matter density. 

Coming back to $w_\eta(z)$, our results could be compared with a few recent studies where the dark energy parameter $w_\Lambda$ of the $\Lambda$CDM model has been reconstructed from minimal extensions of such a basic framework. In particular, \cite{Adame2025} have adopted the standard CPL \cite{Chevallier2001,Linder2003} parametrization $w_\Lambda(a)=w_0+w_a\, (1-a)$ in terms of the scale factor $a$ and fitted BAO+SN data for $(w_0,w_a)$. The result is reported in Figure \ref{fig|EoS} as a cyan line and tends to imply a dark energy strongly decreasing toward the past, so violating the null energy condition for $z\gtrsim 0.5$.
The same analysis based on the CPL parametrization has been refined by \cite{Upala2024} by fitting BAO+SN+SGR data. Their outcome is shown in red, and implies a $w_\Lambda$ mildly decreasing from the present values toward the past, with $w_\Lambda\approx -1$ at $z\gtrsim 1$. Finally, the same authors \cite{Upala2024} have also fitted the BAO+SN+SGR data with a parametrized modification of the expansion history inspired by modified gravity theories, and derived the constraints on $w_\Lambda(z)$ reported in yellow.
Their results suggest a $w_\Lambda(z)$ increasing with redshift, with an overall behavior qualitatively similar to the $\eta$CDM model. This is not surprising, since their parametrization can empirically describe interacting dark energy models \cite[see for instance][]{vanderWesthuizen:2023hcl}, which formally may resemble $\eta$CDM.

All in all, these estimates concerning $w_\Lambda(z)$ are considerably uncertain and critically dependent on the adopted empirical parametrization of the dark energy equation of state and/or of the expansion history, so drawing any strong conclusions is premature. However, the take-home message from the $\eta$CDM model is that the equation of state parameter for the dark energy may be not so fundamental, in that it could just reflect a modified evolution in the expansion history due to an emergent phenomenon.

\subsection{Linder's diagnostic as a crucial probe of \texorpdfstring{$\eta$CDM}{etaCDM}? }\label{sec|cojoin}

\begin{figure}[t!]
    \centering
    \includegraphics[width=0.7\textwidth]{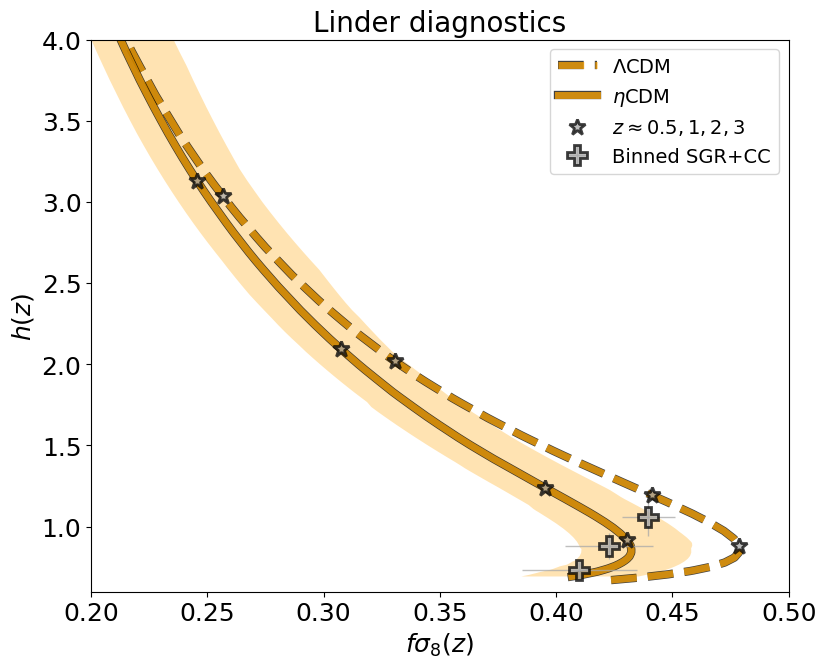}
    \caption{Linder's diagnostic plot of the Hubble rate vs. the structure growth rate. Orange is the result for the $\eta$CDM model from our analysis (solid line is the mean and shaded area is the $1\sigma$ confidence interval). Dashed line refers to the standard $\Lambda$CDM model with parameter set from \texttt{Planck} \cite{Planck2020}. Stars on both the solid and dashed lines highlight the location of the redshifts $z\approx 0.5$, $1$, $2$ and $3$ along the curves from bottom to top. The plus symbols show the CC \cite{Moresco2024} and SGR \cite{Nesseris2017} data binned in common redshift intervals $z<0.2$, $0.2<z<0.6$, and $0.6<z<1.2$ that contain almost the same number of SGR data points.}
    \label{fig|Linderplot}
\end{figure}

As mentioned in Section \ref{sec|intro} it is not easy to identify an observational probe that can clearly distinguish $\Lambda$CDM from any alternative model. This is because, admittedly, the standard scenario fits remarkably well to the cosmological data over a very extended range of spatial and temporal scales, so that any reasonable alternative model should have very similar phenomenological outcomes. Here we propose that a specifically designed observable to test the $\eta$CDM model and possibly to allow selecting or rejecting it against the $\Lambda$CDM framework is constituted by the Linder's diagnostics \cite{Linder:2016xer}. The latter is just a plot of the conjoined background expansion and structure growth rates $H(z)$ vs. $f\sigma_8(z)$, as illustrated in Figure \ref{fig|Linderplot}.

The result for the $\eta$CDM model from our analysis is reported as a solid line with shaded area (median and $1\sigma$ confidence intervals), and compared with the expectation from $\Lambda$CDM \cite{Planck2020} in yellow dashed. For both models, stars highlight the location of redshifts $z\approx 0$, $1$, $2$, $3$ along the curves from bottom to top. Differences between the two models are negligible at high $z$ while they progressively develop toward lower redshift where the effects of the noise associated to structure formation kicks in. In particular, the $\eta$CDM model features a less prominent bump or wiggle at $z\lesssim 0.5-1$. We also report as plus symbols the $f\sigma_8(z)$ data from galaxy surveys \cite{Nesseris2017} and $h(z)$ measurements from cosmic chronometers \cite{Moresco2024}, binned in redshift intervals which contain an almost equal numbers of the former data points (which are less numerous and extend to appreciably lower $z$). We stress that the error bars are just the variance on the stacked mean but do not reflect systematic uncertainties, especially for cosmic chronometers data.

Overall the $\eta$CDM model reproduces very well the binned data, while the $\Lambda$CDM would require $\sigma_8$ to be appreciably lower than the value $\approx 0.81$ expected from CMB. The latter is a manifestation of the well-known $f\sigma_8$ tension suffered by standard $\Lambda$CDM model \cite{Nesseris2017}, which is instead absent in our $\eta$CDM framework. Apart from this general statement, we caveat that uncertainties are still too large and the redshift coverage too narrow in the data to draw definite conclusions. However, future galaxy survey more precisely pinpointing the $f\sigma_8$ parameter in the redshift range $z\lesssim 1$, and more robust determinations of the Hubble rate by cosmology-independent methods such as cosmic chronometers will be extremely relevant and could possibly allow a robust model selection according to this diagnostic plot. 

Note that other interesting diagnostic could be possibly constituted by redshift drift measurements or a combination of these with cosmic chronometers, as proposed by \cite{Koksbang2021}. However, the Linder's test discussed above has the huge advantage that it can be promptly exploited with current or upcoming data, while precise redshift drift measurements will plausibly require decades to be collected.

\begin{figure}[t!]
    \centering
    \includegraphics[width=\textwidth]{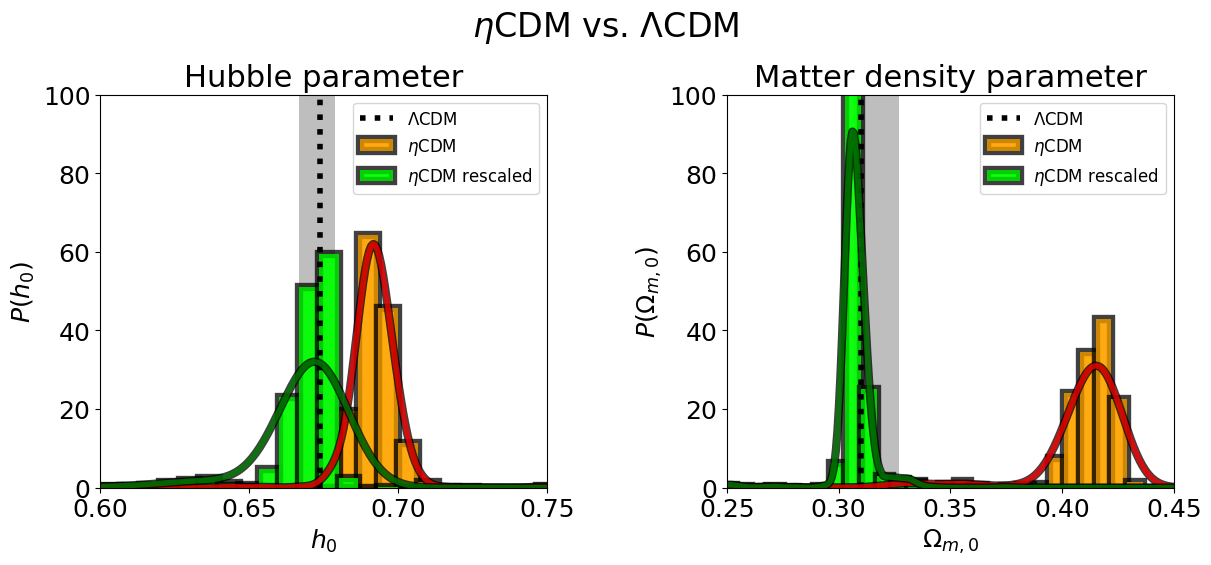}
    \caption{Posterior probability distributions on the Hubble constant $h_0$ (left panel) and on the matter density parameter $\Omega_{m,0}$ (right panel) from our analysis of cosmological data with the $\eta$CDM model. The orange histograms refer to the values at the present time in $\eta$CDM, while the green ones to that measured at the recombination epoch in $\eta$CDM and then scaled to the present according to the $\Lambda$CDM evolution; this is done for a fair comparison with the $\Lambda$CDM results based on \texttt{Planck} CMB data \cite{Planck2020}, which are illustrated by the dotted lines with grey shaded area (median and $1\sigma$ confidence intervals). The solid lines display Gaussian density kernel estimates to the histograms.}
    \label{fig|h0_Om0}
\end{figure}

\subsection{\texorpdfstring{$\eta$CDM}{etaCDM} vs. \texorpdfstring{$\Lambda$CDM}{etaCDM}: Hubble rate and matter density}

In Figure \ref{fig|h0_Om0} we illustrate in orange the marginalized posterior distributions of the Hubble constant $h_0$ and the matter density $\Omega_{m,0}$ in the $\eta$CDM model from our analysis. Specifically, the histograms refers to the binned MCMC chains while the solid lines is a Gaussian kernel density estimate to them. 

One may be concerned that with appreciably high values of the Hubble constant $h_0\approx 0.7$ and above all of the matter density $\Omega_{m,0}\approx 0.4$ in $\eta$CDM there could be problems with reproducing the overall CMB spectrum. However, these values could be misleading, since the evolution of $h(z)$ and $\Omega_m(z)$ in the $\eta$CDM model considerably differ from the standard $\Lambda$CDM scenario after recombination during the era of structure growth (see Figure \ref{fig|evo}). What is really important to reproduce the CMB spectrum are not the values at the present, but those at the recombination epoch $z_\star\approx 1100$. To check that the $\eta$CDM model is safe in this respect, we computed $h(z_\star)$ and $\Omega_{m}(z_\star)$ in $\eta$CDM, and then (just because it is customary to report the values at present time), we evolve them back to the present according to the $\Lambda$CDM model. 
Note that in evolving $h$ and $\Omega_m$ to the recombination redshift in $\eta$CDM we have taken into account the full MCMC chains, hence the correlations between cosmological and noise parameters. In particular, the degeneracy among $h$ vs. $\Omega_m$ vs. $\alpha$ causes the dispersion in $h(z_\star)$ to be slightly larger than that in $h_0$ and the dispersion in $\Omega_m(z_\star)$ to be appreciably smaller than that in $\Omega_{m,0}$. 

The net results are the green histograms and lines (referred in the caption as $\eta$CDM rescaled), that can be directly compared with the bestfit $\Lambda$CDM values from \texttt{Planck} \cite{Planck2020} reported in the Figure as dotted lines with grey shaded areas (median and $1\sigma$ confidence interval). In other words, if the green histograms and the dotted line agree, it means that the recombination values of the Hubble rate $h$ and matter density $\Omega_m$ (hence also the physical density $\rho_m\propto \Omega_m\, h^2$) in $\eta$CDM and $\Lambda$CDM are consistent. This is indeed the case, testifying that no big issue with the full CMB power spectrum is expected. A full analysis of the latter in the $\eta$CDM framework will be pursued in the next future.

The argument can be reversed to better understand why in $\eta$CDM a value of $\Omega_{m,0}\approx 0.4$ higher than in $\Lambda$CDM is found. In fact, we have seen that the matter component in the $\eta$CDM model features a non-trivial equation of state, decreasing from values $w_m\approx 0$ at early times to $w_m<0$ at late times. This means that the dilution of matter by the expansion $\rho_m\propto a^{-3\,(1+w_m)}$ is reduced relative to the standard scenario that has $w_m=0$ hence $\rho_m\propto a^{-3}$ at any epoch. Therefore starting from similar values of $\Omega_m$ at the recombination epoch required to fit CMB data, it is somewhat expected that at present $\Omega_{m,0}$ should be higher in $\eta$CDM than in $\Lambda$CDM.


\begin{figure}[t!]
    \centering
    \includegraphics[width=\textwidth]{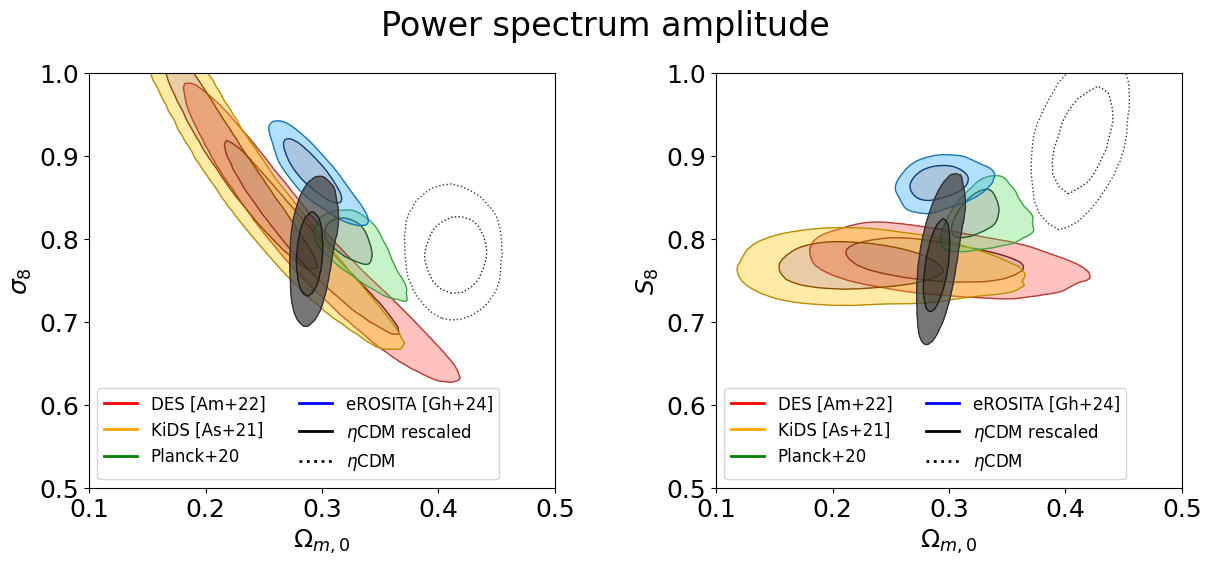}
    \caption{Constraints on the linear matter power spectrum amplitude in terms of the parameter $\sigma_8$ (left panel) or $S_8\equiv \sigma_8\,(\Omega_{m,0}/0.3)^{1/2}$ (right panel) vs. the matter density $\Omega_{m,0}$. Colored contours report the $1\sigma-2\sigma$ confidence intervals from cosmic shear measurements by \texttt{KiDS} \cite{Asgari2021} (orange) and \texttt{DES} \cite{Amon2022} (red), from cluster statistics by \texttt{eROSITA} \cite{Ghirardini2024} (blue), from CMB by \texttt{Planck} \cite{Planck2020} (green), and from our analysis with the $\eta$CDM model (black) when $\Omega_{m,0}$ has been rescaled from the recombination value to the present with the $\Lambda$CDM evolution for fair comparison (for reference, the dotted contours refers to the present values in $\eta$CDM).}
    \label{fig|sigma8_S8}
\end{figure}

\subsection{Broad comparison with other cosmological datasets}\label{sec|otherdata}

In Figure \ref{fig|sigma8_S8} we illustrate as colored shaded areas ($1-2\sigma$ confidence intervals) the constraints in the $\sigma_8$ vs. $\Omega_{m,0}$ and $S_8\equiv \sigma_8\,  (\Omega_{m,0}/0.3)^{1/2}$ vs. $\Omega_{m,0}$
planes from various cosmological probes not employed in this work: cosmic shear measurements from \texttt{DES} \cite{Amon2022} and \texttt{KiDS} \cite{Asgari2021}, full CMB temperature and polarization spectra from \texttt{Planck} \cite{Planck2020}, and cluster statistics from \texttt{eROSITA} \cite{Ghirardini2024}. All these constraints  have been inferred in the standard $\Lambda$CDM scenario and have been shown to stay put in minimal extensions of it. 
From the diagrams one can clearly spot the so called $S_8$ tension of the $\Lambda$CDM model, in that the cosmic shear measurements tend to prefer appreciably lower values of $S_8$ with respect to CMB and cluster statistics.

The outcomes of the $\eta$CDM model from our joint analysis of BAO+SN+SGR+CC data are reported as a grey shaded area. Here $\Omega_{m,0}$ has been rescaled to the present time as discussed in the previous Section, to allow for a fair comparison with the constraints from the other cosmological datasets that are based on the $\Lambda$CDM model (the face value contours of $\eta$CDM would be given by the dotted lines). There is an overall good agreement of the $\eta$CDM constraints from this work with the other datasets reported in Figure \ref{fig|sigma8_S8}. We stress, however, that a full fitting of cosmic shear data, CMB power spectra in temperature and polarization and cluster mass function with the $\eta$CDM model would be required to quantitatively confirm this broad agreement.
This is non-trivial and goes beyond the scope of the present paper, hence it will be pursued in forthcoming works.


\section{Summary and conclusions}\label{sec|summary}

The $\eta$CDM is a new cosmological model by \cite{Lapi2023} aimed to cure at once some drawbacks of the standard $\Lambda$CDM scenario, such as the violation of the cosmological principle at late-times due to structure formation, the origin of the cosmic acceleration, and the cosmic tensions. To this purpose, the model adopts a statistical perspective envisaging a stochastic evolution of large-scale patches in the Universe with typical sizes $10-50\, h^{-1}$ Mpc, which is meant to describe the complex gravitational processes leading to the formation of the cosmic web. The stochasticity among different patches is technically rendered via the diverse realizations of a multiplicative noise term (‘a little ado’) in the cosmological equations, and the overall background evolution of the Universe
is then operationally defined as an average over the patch ensemble (see Section \ref{sec|basics}). In \cite{Lapi2023} we have shown that in $\eta$CDM model an accelerated expansion is naturally obtained at late cosmic time without the need for any postulated dark energy component, since as structure formation proceeds in the Universe an increasing amount of its volume is statistically occupied by low-density regions (i.e., voids) featuring an enhanced expansion rate.

In this work we have provided new theoretical perspectives on the $\eta$CDM model, and we have performed further observational tests against the most recent cosmological datasets. Specifically, our main findings are as follows:

\begin{itemize}

        \item We have shown that the ensemble-averaged evolution of the Universe in $\eta$CDM can be rendered in terms of a spatially flat cosmology and of an emergent dark energy component with a time-dependent equation of state (see Section \ref{sec|emergentDE}), originating the present cosmic acceleration. The effects of the noise can be viewed as transforming a curvature-like term in the early Universe into an emergent dark energy at late times. For the mechanism to work it is required that the ensemble-averaged spatial curvature in the early Universe is negative, albeit it can be arbitrarily small.

       \item We have highlighted that in $\eta$CDM also the matter density features a a non-standard evolution at late times, that may be described in terms of a progressively decreasing equation of state parameter $w_m<0$. Such a behavior often occurs in models with strong coupling in the dark sector, as $\eta$CDM to some extent is: dark energy emerges from structure formation hence depends on the evolution of the matter content, which is in turn affected by the late-time acceleration induced by the emergent dark energy (see Section \ref{sec|emergentDE}). 

        \item We have provided handy expressions for the cosmographic parameters of the $\eta$CDM model, suitable for a quick implementation in the analysis of future datasets (see Section \ref{sec|cosmography}).

        \item We have shown that no coincidence issue is present in the $\eta$CDM model, since at variance with the standard scenario the ensemble-averaged values of the cosmological quantities will stay very close to the present one for a infinite amount of time in the future (see Section \ref{sec|coincidence}).

        \item We have compared the ensemble-averaged evolution of the $\eta$CDM framework to the reference $\Lambda$CDM model, finding that they feature very similar behaviors in terms if the expansion rates, cosmological timescales, cosmographic parameters, distance scales, energy densities, etc. (see Section \ref{sec|evo} and Figure \ref{fig|evo}). The most relevant differences can be spotted in the evolution of the energy density parameters of the matter and of the emergent dark energy (and correspondingly in their equations of state) and in the runs of the jerk and snap parameters at intermediate to high redshift. These could be possibly tested with \texttt{Euclid} via weak lensing and clustering analysis, and with future surveys like those planned on the \texttt{Nancy Grace Roman Space Telescope} via searches for transient objects (e.g., SNe). 

        \item We have tested the $\eta$CDM model against the most recent SN I$a$, BAO and structure growth rate datasets (including the \texttt{Pantheon+} and \texttt{DESI} latest releases) via a standard Bayesian analysis framework (see Section \ref{sec|datasets}), finding a very good agreement (see Section \ref{sec|results}, Figures \ref{fig|MCMC}-\ref{fig|fits} and Table \ref{tab|results}).

        \item We have demonstrated that in $\eta$CDM both the $H_0$ and the $f \sigma_8$ tensions are substantially alleviated, if not solved at all (see Section \ref{sec|results}, Figure \ref{fig|MCMC} and Table \ref{tab|results}). Specifically, the $H_0$ tension is alleviated since, adopting an inverse ladder approach (which is natural since $\eta$CDM is indistinguishable from $\Lambda$CDM at high redshift when structure formation is still in its early and linear stages), values of $h_0\approx 0.7$ could decently fit both BAO and type-I$a$ SN data. Moreover, the $f\sigma_8$ tension is solved since the growth factor of the $\eta$CDM model has a different evolution and allows to fit the structure growth rate data with an appreciably higher $\sigma_8\approx 0.8$ than in $\Lambda$CDM, so as to be consistent with CMB expectations.

        \item We have discussed the evolution of the equation of state for the emergent dark energy, that starts from curvature-like values $w_\eta\approx -1/3$ at early times and decreases toward $w_\eta\lesssim -1$ at late times so enforcing an accelerated expansion. We have also compared our findings to various CPL parameterization adopted in recent analysis of cosmological data with the $\Lambda$CDM model, though the large uncertainties in the reconstruction preclude any robust conclusion (see Section \ref{sec|DE} and Figure \ref{fig|EoS}).

        \item We have stressed that the Linder's diagnostic plot of the redshift-dependent Hubble parameter $H(z)$ vs. structure growth rate $f\sigma_8(z)$ could be a powerful probe of the $\eta$CDM model in the near future (see Section \ref{sec|cojoin} and Figure \ref{fig|Linderplot}). Precise estimates of the $f\sigma_8$ parameter in the redshift range $z\lesssim 1$ by upcoming galaxy surveys, and robust determination of the redshift-dependent Hubble rate by cosmology-independent methods such as cosmic chronometers will be extremely relevant for quantitatively contrasting and distinguishing the $\eta$CDM model against the standard scenario. 
    
        \item Although the $\eta$CDM model requires a present matter density and Hubble parameters $\Omega_{m,0}\approx 0.4$ and $h_0\approx 0.7$ higher than in the standard scenario (based on CMB and BAO data), we have pointed out that the corresponding values of $\Omega_m(z)$ and $h(z)$ at high-redshift agree with those in $\Lambda$CDM, thus ensuring broad consistency with CMB power spectra, cosmic shear surveys and galaxy cluster statistics (see Section \ref{sec|otherdata} and Figures \ref{fig|h0_Om0} and \ref{fig|sigma8_S8}). An in-depth analysis of these datasets is however non-trivial and deferred to future works.     
         
  \end{itemize}

As a concluding remark, we would like to stress that the foundational hypothesis (i.e., a stochastic perspective in place of the deterministic view based on the canonical cosmological principle) and the main outcome (i.e., dark energy as an emergent phenomenon instead of a cosmological constant or dynamical field) of this work could be anyway helpful for future studies aimed at revisiting the standard cosmological framework, independently of the details in the specific model considered here. As to the stochastic perspective, could it be useful to address some further `anomalies' of the standard framework (e.g., \cite{Peebles2022}) like those related to considerable cosmic bulk flows on large scales (see \cite{Tully2023,Hoffman2024}) and/or to the presence of supergiant structures in the Universe (see \cite{Lopez2022,Lopez2024,Boehringer2025})? As to the emergent dark energy, could one relate it to some form of gravitational energy associated to virialized structures/voids or geometric curvature (e.g., \cite{Racz2017,Roukema2018,Cacciatori2024})? All these would constitute extremely relevant questions to be addressed in the near future for paving the way toward an improved model of physical cosmology.

\section*{Acknowledgements}

We thank the referee for the interest in our paper and for illuminating suggestions. We warmly thank Francesco Benetti, Alessandro Bressan, Stefano Liberati, and Tommaso Ronconi for helpful discussions. This work was partially funded from the projects: ``Data Science methods for MultiMessenger Astrophysics \& Multi-Survey Cosmology'' funded by the Italian Ministry of University and Research, Programmazione triennale 2021/2023 (DM n.2503 dd. 9 December 2019), Programma Congiunto Scuole; EU H2020-MSCA-ITN-2019 n. 860744 \textit{BiD4BESt: Big Data applications for black hole Evolution STudies}; Italian Research Center on High Performance Computing Big Data and Quantum Computing (ICSC), project funded by European Union - NextGenerationEU - and National Recovery and Resilience Plan (NRRP) - Mission 4 Component 2 within the activities of Spoke 3 (Astrophysics and Cosmos Observations);  European Union - NextGenerationEU under the PRIN MUR 2022 project n. 20224JR28W "Charting unexplored avenues in Dark Matter"; INAF Large Grant 2022 funding scheme with the project "MeerKAT and LOFAR Team up: a Unique Radio Window on Galaxy/AGN co-Evolution; INAF GO-GTO Normal 2023 funding scheme with the project "Serendipitous H-ATLAS-fields Observations of Radio Extragalactic Sources (SHORES)". LB acknowledges financial support from the German Excellence Strategy via the Heidelberg Cluster of Excellence (EXC 2181 - 390900948) STRUCTURES. AL dedicates this work to his father Giovanni.

\bibliographystyle{JHEP}
\bibliography{bibliografia}

\begin{appendix}

\section{Appendix: \texorpdfstring{$\eta$CDM}{etaCDM} and cosmological spatial averaging} \label{app|spave}

In this Appendix we review the issue of spatial averaging in Newtonian cosmology, as developed by many literature studies focused on deterministic backreaction (e.g., \cite{Buchert1997,Buchert2000}), to establish a connection with the $\eta$CDM model. Specifically, we aim to clarify the origin, and quantitatively assess the scale-dependent strength of the stochasticity in the mass-energy conservation equation that is at the heart of the $\eta$CDM framework.

Our discussion is based on the well-known assumptions that the large-scale dynamics of the Universe can be approximately described in Newtonian gravity (as adopted in cosmological numerical simulations; e.g., for a review see \cite{Vogelsberger2020}), and that the evolution of the various matter and energy components can be treated in the fluid approximation. In this vein, we start by writing the continuity, Euler and Poisson equation for a (pressureless) cosmological fluid with density field $\rho({\bf r},t)$ and velocity field ${\bf u}({\bf r},t)$ as:
\begin{equation}
\left\{
\begin{aligned}
& \partial_t\,\rho + \nabla_{\bf r}\, (\rho\, {\bf u}) = 0~,\\
&\\
& \partial_t\, {\bf u} + ({\bf u} \cdot \nabla_{\bf r})\, {\bf u} = -\nabla_{\bf r}\,\phi~,\\
&\\
& \nabla_{\bf r}^2\,\phi = 4\pi G\, \rho~.
\end{aligned}
\right.
\end{equation}
We switch to the comoving coordinate ${\bf x}$ defined via ${\bf r} = a(t)\, {\bf x}$, with $a(t)$ being the scale factor. We obtain ${\bf u} = \dot {\bf r} = \dot a\, {\bf x}+a\, \dot {\bf x}$, where the last addendum is just the peculiar velocity. In the new coordinates $\nabla_{\bf r} = (1/a)\, \nabla_{\bf x}$ and $\partial_t|_{{\bf r}=\rm const.} = \partial_t|_{{\bf x}=\rm const.}-H\, ({\bf x}\cdot \nabla_{\bf x})$ hold, with $H\equiv \dot a/a$ representing the expansion rate. The continuity equation can be rewritten as
\begin{equation}\label{eq|fieldrho}
\dot\rho + 3 H\, \rho = -\rho\, \theta~,
\end{equation}
where we have defined the comoving peculiar velocity divergence field $\theta \equiv \nabla_{\bf x}\cdot \dot {\bf x}$ and the total derivative along the fluid flow ${\rm d}_t = \dot{ } \equiv \partial_t+\dot {\bf x}\cdot \nabla_{\bf x}$. In addition, the Euler equation can be recast in the form
\begin{equation}
\partial_t\, \dot {\bf x} + (\dot {\bf x}\cdot\nabla_{\bf x})\, \dot {\bf x}+ 2 H\, \dot {\bf x} = - \frac{1}{a^2}\, \Phi~,
\end{equation}
where $\Phi\equiv \phi+ a\, \ddot a\, {\bf x}^2/2$ is the peculiar gravitational potential. Finally, the Poisson equation turns into
\begin{equation}
\frac{1}{a^2}\, \nabla_{\bf x}^2 \Phi  = \left(4\pi G\,\rho+3\, \frac{\ddot a}{a}\right)~,
\end{equation}
where on the r.h.s. one can recognize terms that appear in the standard cosmological acceleration equation.
It is convenient to take the divergence of the Euler equation, insert in it the Poisson one, and use the vector decomposition $\nabla_{\bf x}\cdot [(\dot {\bf x}\cdot \nabla_{\bf x})\, \dot {\bf x}] = (\dot {\bf x}\cdot \nabla_{\bf x})\,\theta+ \frac{1}{3}\, \theta^2+2\,(\sigma^2-\omega^2)$, where $\sigma\equiv \sqrt{\sigma_{ij}\sigma^{ij}/2}$ and $\omega\equiv \sqrt{\omega_{ij}\omega^{ij}/2}$ are the magnitudes of the shear $\sigma_{ij}$ and rotation $\omega_{ij}$ tensors, respectively. Putting all together, one gets
\begin{equation}\label{eq|fieldtheta}
\dot \theta + 2H\,\theta + \frac{\theta^2}{3}+2\,(\sigma^2-\omega^2) = -\left(3\frac{\ddot a}{a}+4\pi G\rho\right)~.
\end{equation}

The above field Equations (\ref{eq|fieldrho}) and (\ref{eq|fieldtheta}) are supposed to hold at any given spatial locations, but we now proceed to take spatial averages over a comoving volume $V(t)\equiv \int{\rm d^3}{\bf x}$. 
First of all, one can notice that  
\begin{equation}
\begin{aligned}
\frac{\dot V}{V} = \frac{1}{V}\, {\rm d}_t \int_{V(t)}{\rm d^3}{\bf x}  &= \frac{1}{V}\, {\rm d}_t \int_{V(t_0)}{\rm d^3}{\bf x_0}\, |J|  =\frac{1}{V}\,\int_{V(t_0)}{\rm d^3}{\bf x_0}\, \dot{|J|}= \\
&\\
&= \frac{1}{V}\,\int_{V(t_0)}{\rm d^3}{\bf x_0}\, |J|\, \theta = \frac{1}{V}\,\int_{V(t)}{\rm d^3}{\bf x}\, \theta = \langle\theta\rangle~,
\end{aligned}
\end{equation}
where we have defined the Jacobian $J \equiv \nabla_{\bf x_0} {\bf x}$ and used the fact that $\dot{J} = \nabla_{\bf x_0} \dot {\bf x} = (\nabla_{\bf x_0} \dot {\bf x}/\nabla_{\bf x_0} {\bf x})\, \nabla_{\bf x_0} {\bf x}= \nabla_{\bf x} \dot {\bf x}\, \nabla_{\bf x_0} {\bf x} = \theta\, J$ holds (symbolic calculus employed here). Then the spatial-average and time-derivative operations of a generic field $\mathcal{F}$ are seen not to commute since 
\begin{equation}
\overbigdot{\langle \mathcal{F}\rangle} = {\rm d}_t \left(\frac{1}{V}\, \int_{V(t)}{\rm d}^3 {\bf x}\, \mathcal{F}\right) = -\frac{\dot V}{V}\, \langle\mathcal{F}\rangle+\frac{1}{V}\, \int_{V(t_0)}{\rm d}^3{\bf x_0}\, {\rm d}_t(\mathcal{F}\, J) = -\langle\theta\rangle\, \langle\mathcal{F}\rangle + \langle\dot{\mathcal{F}}\rangle+\langle\mathcal{F}\theta\rangle~.
\end{equation}
Applying the averaging procedure to Equations (\ref{eq|fieldrho}) and (\ref{eq|fieldtheta}) we obtain
\begin{equation}\label{eq|eqave}
\left\{
\begin{aligned}
& \overbigdot{\langle\theta\rangle} + 2H\, \langle\theta\rangle+ \frac{\langle\theta\rangle^2}{3} = \frac{2}{3}\, \sigma_\theta^2+2\,(\langle\omega^2\rangle-\langle\sigma^2\rangle) - \left(3\frac{\ddot a}{a}+4\pi G\,\langle\rho\rangle\right)~,\\
&\\
& \overbigdot{\langle\rho\rangle}+ 3H\, \langle\rho\rangle = -\langle\rho\rangle \langle\theta\rangle~,
\end{aligned}
\right.
\end{equation}
where $\sigma_\theta^2\equiv \langle\theta^2\rangle-\langle\theta\rangle^2$ is the variance of the $\theta$  field.
Up to now, the background $H$ and $a$ have not been specified: the natural choice is to require that they represent the average scale factor and expansion rate of the spatial region; this automatically will imply that the peculiar velocity divergence $\langle\theta\rangle=0$ is null since the average expansion of the volume will be fully described by $a$ and $H$. This is also confirmed by observations of local cosmic flows, when averaging over sufficiently large scales $\gtrsim 10\, h^{-1}$ Mpc (see \cite{Hoffman2024}). Putting $\langle\theta\rangle=0$ in Equation (\ref{eq|eqave}) we thus obtain
\begin{equation}
\left\{
\begin{aligned}
& \frac{\ddot a}{a} = -\frac{4\pi G}{3}\,\langle\rho\rangle + \frac{\mathcal{Q}}{3}~,\\
&\\
& \overbigdot{\langle\rho\rangle}+ 3H\, \langle\rho\rangle = 0~,
\end{aligned}
\right.
\end{equation}
in terms of the `backreaction' variable $\mathcal{Q}\equiv \frac{2}{3}\, \langle\theta^2\rangle+2\,(\langle\omega^2\rangle-\langle\sigma^2\rangle)$. Integrating in time the first equation and using the second, we arrive at the spatially-averaged Friedmann equation
\begin{equation}
\frac{\dot{a}^2+k}{a^2} = \frac{8\pi G}{3}\,\langle\rho\rangle + \frac{2}{3\, a^2}\int{\rm d}a\, a\, \mathcal{Q}~.
\end{equation}
Finally, it is convenient to redefine the density in terms of an effective matter source $\langle\rho_{\rm eff}\rangle\equiv \langle\rho\rangle + \frac{1}{4\pi G a^2}\, \int{\rm d}a\, a\, \mathcal{Q}$, that leads to the system
\begin{equation}\label{eq|backreaction}
\left\{
\begin{aligned}
& \left(\frac{\dot a}{a}\right)^2 = \frac{8\pi G}{3}\,\langle\rho_{\rm eff}\rangle - \frac{k}{a^2}~,\\
&\\
& \overbigdot{\langle\rho_{\rm eff}\rangle}+ 3H\, \langle\rho_{\rm eff}\rangle = \frac{H\mathcal{Q}}{4\pi G}\, \left(1+\frac{1}{a^2\, \mathcal{Q}}\, \int{\rm d}a\, a\, \mathcal{Q}\right)~.
\end{aligned}
\right.
\end{equation}

One may wonder if the constant $k$ is somehow related to the spatially-averaged scalar curvature of the region. A general relativistic treatment of the spatial averaging procedure carried out by \cite{Buchert2000,Buchert2008,Buchert2012,Buchert2013} shows that
\begin{equation}\label{eq|kappaspave}
\frac{k}{a^2} = \frac{\langle \mathcal{R}\rangle+\mathcal{Q}}{6}+\frac{2}{3\, a^2}\int{\rm d}a\, a\, \mathcal{Q}~,
\end{equation}
holds in terms of the spatially-averaged three-dimensional Ricci scalar $\langle\mathcal{R}\rangle$. This is because inhomogeneities in the matter distribution may trigger the developments of spatially-averaged curvature also in regions that were originally close to flatness with small $k$ (see also \cite{Kolb2011,Wiltshire2007}). All in all, the constant $k$ must be thought as an initial condition related to the spatial curvature pertaining to the patch under consideration at early times (when $\mathcal{Q}\approx 0$), but this direct link with the scalar curvature is lost as inhomogeneities/anisotropies develop at late times.

Note that the backreaction term $\mathcal{Q}$ in Equations (\ref{eq|backreaction}) and (\ref{eq|kappaspave}) has been subject to a longstanding debate in the community since the early conjecture that under certain conditions it could possibly affect the expansion rate of (a region in) the Universe and led to an overall acceleration \cite{Kolb2011,Buchert2012}. On the one hand, it has been pointed out that the impact of (deterministic) backreaction 
depends on the size of the region, which rapidly decreases for scales larger than the largest inhomogeneity \cite{Buchert1997,Kaiser2017,Buchert2018}, as also indicated by numerical simulations in a general relativistic setting \cite{Adamek2019,Macpherson2019}. However, these conclusions strongly rely on the assumption that it is possible to define a global background metric (or a global rescaling of it to different scales) and a deterministic backreaction term. This is not at all guaranteed at late times on scales $\lesssim 50\, h^{-1}$ Mpc pertaining to the cosmic web and relevant for several cosmological observables. There the different evolutions of the various patches due to local inhomogeneities/anisotropies, cosmic flows, tidal forces, sample variance, and many complex gravitational processes cannot be neglected and call for a statistical approach to cosmology as pursued in the main text.

\begin{figure}[t!]
    \centering
    \includegraphics[width=1.\textwidth]{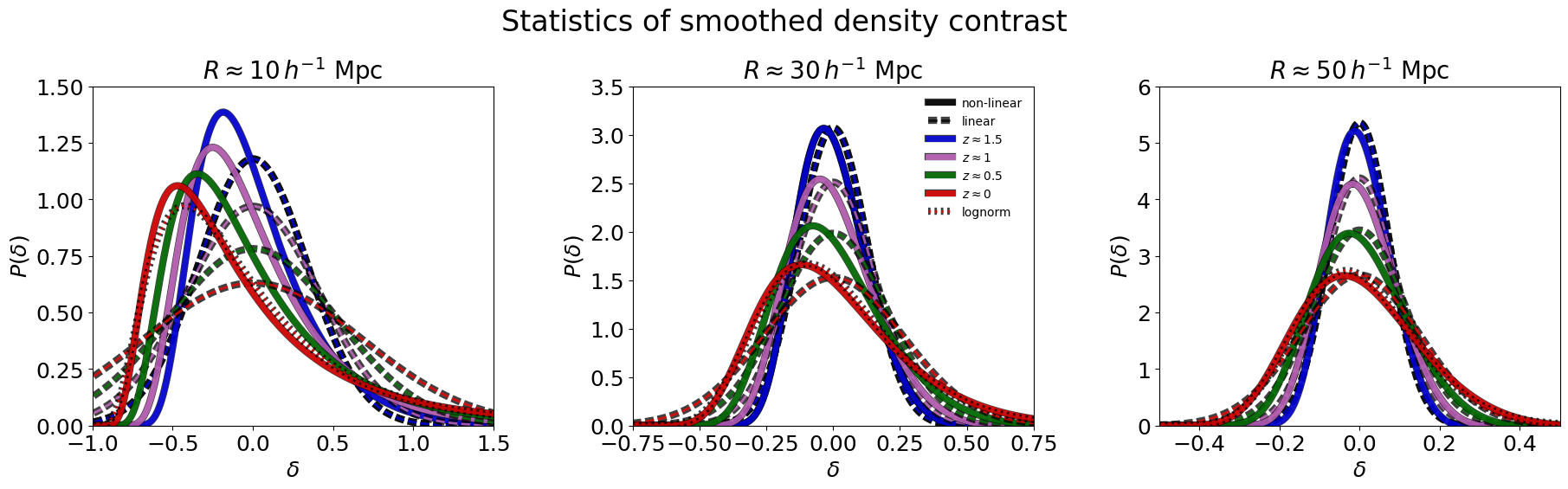}
    \caption{Statistics of the overdensity field smoothed on different comoving spatial scales $R\approx 10\, h^{-1}$ Mpc (left), $30\, h^{-1}$ Mpc (middle) and $50\, h^{-1}$ Mpc (right) at various redshifts $z\approx 1.5$ (blue), $1$ (magenta), $0.5$ (green) and $0$ (red). Dashed line illustrate the outcomes from linear evolution (Gaussian shapes) and solid lines the true, non-linear distributions (approximately lognormal shapes). The red dotted line is an exactly lognormal distribution with the same mean and variance of the true one at $z\approx 0$, to highlight the slight deviation of the true distribution from the lognormal behavior. The non-linear statistics is computed via the fitting formula to numerical simulations by \cite{Repp2018}.}
    \label{fig|deltastats}
\end{figure}

To make contact between Equations (\ref{eq|backreaction}) and the $\eta$CDM framework, we can (naively) proceed as follows. First of all, since the constant curvature parameter $k$ is typically small on the large scales considered here, we can use the Friedmann equation to approximately write $H^2\propto \langle\rho_{\rm eff}\rangle$ in the mass-energy evolution equation, to obtain
\begin{equation}\label{eq|noise_expl}
\overbigdot{\langle\rho_{\rm eff}\rangle}+ 3H\, \langle\rho_{\rm eff}\rangle = \frac{2}{3}\, \langle\rho_{\rm eff}\rangle\, H^{-1}\, \mathcal{Q}\, \left(1+\frac{1}{a^2\, \mathcal{Q}}\, \int{\rm d}a\, a\, \mathcal{Q}\right)~.
\end{equation}
Then we can have an hint on the stochastic nature of the variable $\mathcal{Q}\sim \langle\theta^2\rangle$ from the following argument. It is well known from theoretical considerations and numerical simulations that the density contrast field $1+\langle\delta\rangle\propto \langle\rho\rangle$ smoothed on a given large spatial scale (e.g., tens of Mpc) follows a Gaussian distribution at early times: this reflects the initial conditions that are almost preserved during linear evolution. However, at late times its distribution shifts toward an approximately lognormal shape (see \cite{Coles1991,Repp2017,Repp2018}); this is because complex gravitational processes associated to nonlinear structure formation tend to aggregate matter in the filaments/nodes of the cosmic web, while subtracting it from other regions that will turn out to be the voids. In Figure \ref{fig|deltastats} one can visualize the late-time evolution for the statistics of the density contrast smoothed on different comoving scales $R\approx 10-30-50\, h^{-1}$ Mpc; this has been computed via the precise fitting formula to numerical simulations by \cite{Repp2018}. It is evident, especially for scales $\lesssim 30\, h^{-1}$ Mpc that the non-linear evolution associated to structure formation kicks in at $z\lesssim 1$ to change appreciably the initial overdensity distribution, converting it from a Gaussian to a closely lognormal shape.

All the above means that sampling different large-scale patches in the late-time Universe would yield a closely lognormal distribution for the related values of $\langle\rho\rangle\propto 1+\langle\delta\rangle$ or, equivalently, a normal distribution for $\log\langle\rho\rangle$. In terms of basic stochastic processes, such a distribution would be naturally created by an ensemble of regions whose spatially-averaged density evolves stochastically in time under a Gaussian white noise $\eta(t)$ or, in other words, for which
${\rm d}_t \log \langle\rho\rangle \propto \overbigdot{\langle \rho\rangle}/\langle\rho\rangle\propto \eta(t)$ applies (see \cite{Risken1996,Paul2013}; also Appendix A in \cite{Lapi2023}). To realize this in Equation (\ref{eq|noise_expl}), it is natural to assume the scaling
\begin{equation}
\mathcal{Q}(t)\approx \langle\theta^2\rangle\propto\langle\theta^2\rangle_\star\, H^{\alpha+1}\, \eta(t)~,
\end{equation}
where $\langle\theta^2\rangle_\star$ is the value of the peculiar divergence velocity field at a reference time, and $\alpha$ is a parameter (expected to be close to $-1$), which describes possible deviations from a pure additive noise behavior. Since in our framework the noise term associated to $\mathcal{Q}\approx\langle\theta^2\rangle\propto \eta(t)$ is meant to describe small deviations from homogeneity/isotropy, the quantity $\frac{1}{a^2}\,\int{\rm d}a\, a \, \mathcal{Q}$ entering Equation (\ref{eq|noise_expl}) and the definition of $\langle\rho_{\rm eff}\rangle$, where the fluctuating variable $\mathcal{Q}$ is averaged over cosmic time, could be considered of sub-leading order. This approximately implies $\langle\rho_{\rm eff}\rangle\simeq \langle\rho\rangle$ and yields the mass-energy evolution equation
\begin{equation}
\overbigdot{\langle\rho\rangle}+ 3H\, \langle\rho\rangle = \zeta\, \langle\rho\rangle\, \left(\frac{H}{H_\star}\right)^{\alpha}\, H_\star^{1/2}\, \eta(t)~,
\end{equation}
where the adimensional parameter $\zeta\propto \langle\theta^2\rangle_\star$ encases all the proportionality constants, and $H_\star$ is an arbitrary reference value of the Hubble parameter which ensures dimensional consistency since the noise $\eta$ has dimension of $1/\sqrt{\rm time}$.
In fact, a stochastic equation of such form is adopted in the $\eta$CDM model. 

Plainly, the strength parameter $\zeta\propto \langle\theta^2\rangle_\star$ is related to the spatially-averaged variance of the divergence of the peculiar velocity field, and will depend on the smoothing scale. Over the size of the whole Universe, one should expect $\zeta\propto \langle\theta^2\rangle\approx 0$, and so there will be no impact of the stochasticity. However, on scales $R\lesssim 50\, h^{-1}$ Mpc associated to the emergence of the cosmic web, that are crucial to many cosmological data, the value of $\zeta\propto \langle\theta^2\rangle$ can be non-negligible. This occurrence will bring about a relevant impact of the stochasticity in the overall late-time cosmological evolution.

To summarize, although in the original $\eta$CDM model by \cite{Lapi2023} the stochasticity in the mass-energy evolution equation was somewhat postulated or at most justified only qualitatively, the above lines of reasoning show that it can be actually understood in the context of spatial averaging in Newtonian cosmology, and its scale-dependent strength $\zeta\propto \langle\theta^2\rangle_\star$ can be quantitatively related to the value of the spatially-averaged variance of the divergence of the peculiar velocity field. It will be extremely interesting to precisely measure the latter from reconstruction of the cosmic velocity pattern.

\section{Growth of linear perturbations in \texorpdfstring{$\eta$CDM}{etaCDM}}\label{app|linpert}

In this Appendix we derive the equation for the evolution of linear perturbations in the $\eta$CDM model. One can start by writing $\langle\rho\rangle\simeq \rho_0\,(1+\delta)$ in Equation (\ref{eq|eqave}) where $\delta$ is a small perturbation in any given patch of background density $\rho_0$.  An important point to recall is that $\theta \equiv \nabla_{\bf x}\cdot \dot {\bf x}$ is the divergence of the peculiar velocity field and hence is also small (same order of $\delta$, see below). This implies that the backreaction variable $\mathcal{Q}\sim \langle\theta^2\rangle$ can be neglected at linear order, hence no stochasticity enters into the game and all the patches (and plainly also the ensemble average) would evolve in the same way; up to the linear regime there is no difference with respect to the standard cosmological evolution. In fact, at zeroth order one gets the standard Friedmann equations for an Einstein de Sitter Universe, while at first order one has
\begin{equation}\label{eq|growth}
\left\{
\begin{aligned}
& \dot\theta+2 H\, \theta= - 4\pi\,G\, \rho_0\, \delta~,\\
&\\
& \dot \delta \simeq - \theta~,
\end{aligned}
\right.
\end{equation}
or equivalently
\begin{equation}
\ddot \delta +2 H\,\dot\delta=4\pi G\, \rho_0\,\delta~.
\end{equation}
Passing to log-conformal time ${\rm d}\ln a = H\, {\rm d}t$ and indicating with a prime the derivatives with respect to $\ln a$, one obtains
\begin{equation}
\delta''+\left(2+\frac{H'}{H}\right)\,\delta'=\frac{3}{2}\,\Omega_m\,\delta~,
\end{equation}
which is Equation (\ref{eq|linpert}) of the main text. As for the smoothed velocity field divergence, from the second of Equations (\ref{eq|growth}) one finds that
\begin{equation}
\theta= - \delta\, H\, f~,
\end{equation}
where $f(a)\equiv \delta'/\delta$ is the growth rate of perturbations in linear theory. 

\begin{figure}[t!]
    \centering
    \includegraphics[width=.7\textwidth]{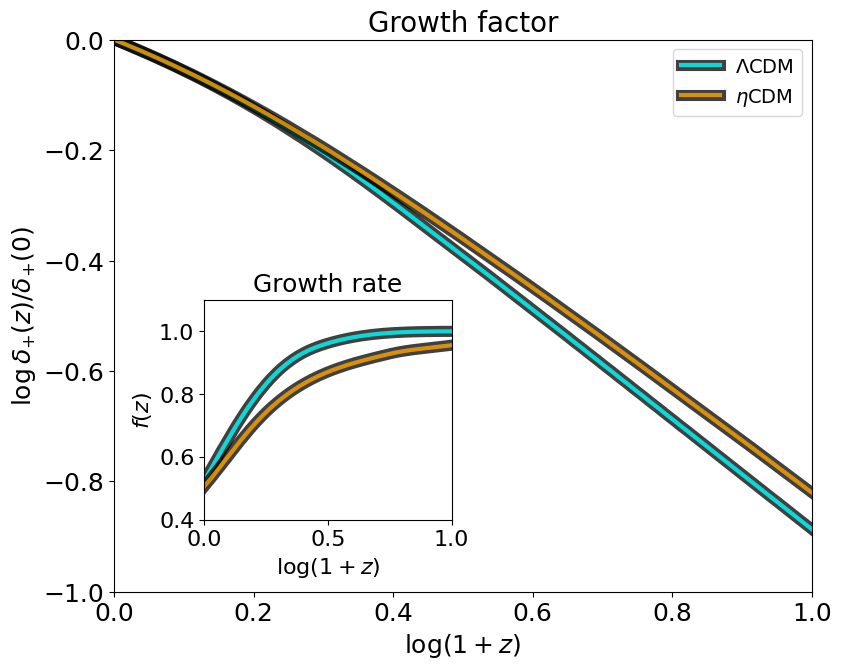}
    \caption{The growing mode solution $\delta_+(z)$ of Equation (\ref{eq|growth}) normalized to the present time as a function of redshift; the inset reports the redshift evolution of the corresponding growth rate $f(a)={\rm d}\ln \delta_+/{\rm d}\ln a$. Cyan solid lines refer to $\Lambda$CDM and orange solid lines to $\eta$CDM. For reference the same sets of cosmological parameters of Figure \ref{fig|evo} have been adopted.}
    \label{fig|growth}
\end{figure}

A subtle point in $\eta$CDM (but also occurring in other scenarios of modified gravity) is that the zeroth order solution in perturbation theory is substantially different from the non-perturbative ensemble-averaged evolution of the Universe analyzed in the main text. This is because, as noted above, up to linear order stochasticity is negligible, so all patches of the Universe (and their ensemble average) would evolve similarly and no relevant effects, including the emergence of dark energy, would occur. However, it is more reasonable to compute the evolution of perturbations over the true non-perturbative ensemble-averaged background of the $\eta$CDM model, which is induced by nonlinear structure formation. Thus when exploiting the equation of linear perturbations derived above, we use as background the ensemble-averaged evolution of the Hubble parameter $\bar H$ and of the matter density parameter $\bar\Omega_m$ found by solving in full Equations (\ref{eq|friednew}).

The growing mode solution $\delta_+(z)$ of Equation (\ref{eq|growth}) normalized to the present time is illustrated in Figure \ref{fig|growth}, while the inset shows the corresponding growth rate $f(a)={\rm d}\ln \delta_+/{\rm d}\ln a$. Cyan solid lines refer to $\Lambda$CDM and orange solid lines to $\eta$CDM. For reference we have assumed the same sets of cosmological parameters of Figure \ref{fig|evo} in the main text. It is seen that the growth of perturbation in $\eta$CDM is slower than in $\Lambda$CDM; this would yield less abundant, but more massive structures at high redshift in the former with respect to the latter. It would be interesting to investigate the consequence of such a slower growth rate in $\eta$CDM on the halo mass function, and check whether it could help explaining the overabundance of massive galaxies at high $z$ as found by \texttt{JWST} surveys (e.g., \cite{Labbe2023,Rashi2024}); we plan to investigate the issue in the near future.

In $\Lambda$CDM it is well known (e.g., \cite{Linder_2003}) that a good approximation to the growth rate is given by $f(a)\simeq \Omega_m^{\gamma}(a)$, where the value $\gamma \approx 0.55$ of the `growth index' is obtained when assuming standard general relativity \cite{Fry:1985zy, Linder:2005in,Wang:1998gt, Gong:2008fh}. In the $\eta$CDM model we have found that the growth rate can be very well approximated as
\begin{equation}
f(a)\simeq \Omega_m^{\gamma(a)}(a) ~,
\end{equation}
in terms of a slowly evolving growth index $\gamma(a)\approx 0.55+0.15\,a$; this could be thought as a late-time deviation from the $\Lambda$CDM value $\gamma\approx 0.55$ applying at high redshift, where the initial conditions to solve Equation (\ref{eq|growth}) have been set. Remarkably, the outcome is in extremely good agreement with the recent claims for a larger value of $\gamma \gtrsim 0.64$ presented in \cite{Nguyen:2023fip}, who quote a highly significant tension with the concordance cosmology. In fact, the deviation of the growth index from the standard value $\gamma \approx 0.55$ is usually interpreted as an evidence in favour of modified gravity scenarios \cite{Linder:2007hg, Gong:2008fh}. However, in the $\eta$CDM models such steepening of the growth index toward low $z$ can be easily explained without invoking any modifications to the fundamental nature of gravity. 

Finally, the redsfhit dependence of the growth index is such that it starts to deviate from the standard value  $\gamma\approx 0.55$ well within the matter-dominated regime, at around  $z\lesssim 1 $. Incidentally, this is around the cosmic epoch where the CMB lensing kernel peaks; in fact, a higher $\gamma$ as found in $\eta$CDM immediately implies a higher CMB lensing amplitude \cite{Nguyen:2023fip}. This is in fact consistent with observations and constitutes the well established, albeit mild, $A_{\rm lens}$-or lensing-anomaly of the $\Lambda$CDM scenario \cite{Planck2020,Addison2024}. All in all, this could provide a hint that the $\eta$CDM is able to solve yet another cosmic tension. Although a clear-cut demonstration for the superior performances of the $\eta$CDM model over the standard scenario is still far away, the above clue adds to the many assessments presented so far in this work to provide a positive outlook for future investigations of our stochastic approach to cosmology.

\end{appendix}

\end{document}